\documentclass[useAMS,usenatbib,usegraphicx]{mn2e}

\usepackage[british]{babel} 
\usepackage{times}
\usepackage{latexsym,amssymb}
\usepackage[fleqn]{amsmath}
\usepackage{graphicx}

\usepackage{longtable}

\usepackage{natbib}
\usepackage{journalnames}
\usepackage{color}
\usepackage{graphicx}
\usepackage{epstopdf}
\usepackage[T1]{fontenc}
\usepackage{aecompl}

\usepackage{caption}
\usepackage{comment}
\usepackage{lipsum}
\usepackage{array}
\usepackage{dcolumn}

\usepackage{amsmath} % or simply amstext
\usepackage{gensymb}  % add degree symbol

\title[The EDGE-CALIFA survey: the influence of galactic rotation on the molecular depletion time]{The EDGE-CALIFA survey: the influence of galactic rotation on the molecular depletion time across the Hubble sequence}

\author[D. Colombo et al.]{%
D. Colombo$^1$\thanks{E-mail: dcolombo@mpifr-bonn.mpg.de}, 
V. Kalinova$^1$,
D. Utomo$^{2,3}$, 
E. Rosolowsky$^4$,
A. D. Bolatto$^5$,
R. C. Levy$^5$,
\newauthor{T. Wong$^6$,
S. F. Sanchez$^7$,
A. K. Leroy$^3$,
E. Ostriker$^8$,
L. Blitz$^2$, 
S. Vogel$^5$,
D. Mast$^9$,}
\newauthor{R. Garc\'ia-Benito$^{10}$, B. Husemann$^{11}$, H. Dannerbauer$^{12,13}$,
L. Ellmeier$^{14}$, Y. Cao$^6$.}
\\
$^1$ Max Planck Institute for Radioastronomy, Auf dem H\"ugel 69, D-53121, Bonn, Germany\\
$^2$ Department of Astronomy, University of California, Berkeley, CA 94720, USA\\
$^3$ Department of Astronomy, The Ohio State University, 140
West 18th Avenue, Columbus, OH 43210, USA\\
$^4$ Department of Physics, University of Alberta, 4-181 CCIS, Edmonton, AB T6G 2E1, Canada\\
$^5$ Department of Astronomy, University of Maryland, College Park, MD 20742, USA\\
$^6$ Department of Astronomy, University of Illinois, Urbana, IL 61801, USA\\
$^7$ Instituto de Astronomi\'a, Universidad Nacional Auton\'oma de Mexico, A.P. 70-264, 04510 M\'exico, D.F., Mexico\\
$^8$ Department of Astrophysical Sciences, Princeton University, Princeton, NJ 08544, USA\\
$^9$ Observatorio Astron\'omico de C\'ordoba, Laprida 854, Observatorio, X5000BGR C\'ordoba, Argentina\\
$^{10}$ Instituto de Astrof\'isica de Andaluc\'ia (CSIC), Glorieta de la Astronom\'ia s/n, Aptdo. 3004, 18080 Granada, Spain\\
$^{11}$ Max-Planck-Institut f\"ur Astronomie, K\"onigstuhl 17, D-69117 Heidelberg, Germany\\
$^{12}$ Instituto de Astrof\'isica de Canarias, E-38205 La Laguna, Tenerife, Spain\\
$^{13}$ Universidad de La Laguna, Dpto. Astrof\'isica, E-38206 La Laguna, Tenerife, Spain\\
$^{14}$ Department of Astrophysics, University of Vienna, T\"urkenschanzstrasse 17, 1180, Vienna, Austria
\\
}
\pagerange{\pageref{firstpage}--\pageref{lastpage}} \pubyear{2017}

\begin{document}

\newcommand{\hi}          {\mbox{\rm H{\small I}}}
\newcommand{\HI}         {\hi}
\newcommand{\hii}         {\mbox{\rm H{\small II}}}
\newcommand{\htwo}        {\mbox{H$_{2}$}}
\newcommand{\jone}        {\mbox{$J=1-0$}}
\newcommand{\jtwo}        {\mbox{$J=2-1$}}
\newcommand{\jthree}        {\mbox{$J=3-2$}}
\newcommand{\jfour}        {\mbox{$J=4-3$}}
\newcommand{\yr}      {\mbox{\rm yr}}
\newcommand{\um}          {$\mu$m}
\newcommand{\ha}          {H$\alpha$}
\newcommand{\hb}          {H$\beta$}
\newcommand{\kmpers}      {\mbox{\rm km~s$^{-1}$}}
\newcommand{\kms} {\kmpers}
\newcommand{\percmcu}     {\mbox{\rm cm$^{-3}$}}
\newcommand{\msun}        {\mbox{\rm M$_\odot$}}
\newcommand{\msol}{\msun}
\newcommand{\msunperpcsq} {\mbox{\rm M$_\odot$~pc$^{-2}$}}
\newcommand{\msunperkpcsq} {\mbox{\rm M$_\odot$~kpc$^{-2}$}}
\newcommand{\msunperpcsqyr} {\mbox{\rm M$_\odot$~pc$^{-2}$~yr$^{-1}$}}
\newcommand{\msunperyr}   {\mbox{\rm M$_\odot$~yr$^{-1}$}}
\newcommand{\msunperpccu} {\mbox{\rm M$_\odot$~pc$^{-3}$}}
\newcommand{\msunperyrkpcsq} {\mbox{\rm M$_\odot$~yr$^{-1}$~kpc$^{-2}$}}
\newcommand{\amsunperkpcsq} {\mbox{\rm M$_\odot$/kpc$^{2}$}}
\newcommand{\amsunperpcsq} {\mbox{\rm M$_\odot$/pc$^{2}$}}
\newcommand{\amsunperyrkpcsq} {\mbox{\rm M$_\odot$/(yr~kpc$^{2}$)}}
\newcommand{\reff}         {\mbox{$R_{\rm e}$}}
\newcommand{\rgal}         {\mbox{$R_{\rm gal}$}}
\newcommand{\xco}         {\mbox{$X_{\rm CO}$}}
\newcommand{\xcot}         {\mbox{$X_{\rm CO,20}$}}
\newcommand{\aco}         {\mbox{$\alpha_{\rm CO}$}}
\newcommand{\xcounits}    {\mbox{\rm cm$^{-2}$(K km s$^{-1}$)$^{-1}$}}
\newcommand{\acounits}  {\mbox{\rm M$_\odot$ (K km s$^{-1}$ pc$^2$)$^{-1}$}}
\newcommand{\Lcounits}  {\mbox{\rm K km s$^{-1}$ pc$^2$}}
\newcommand{\Kkmpers}     {\mbox{\rm K km s$^{-1}$}}
\newcommand{\Kkmperspcsq} {\mbox{\rm K km s$^{-1}$ pc$^2$}}
\newcommand{\co}          {\mbox{$^{12}$CO}}
\newcommand{\cothree}          {\mbox{$^{13}$CO}}
\newcommand{\Ico}         {\mbox{I$_{\rm CO}$}}
\newcommand{\av}          {\mbox{$A_V$}}
\newcommand{\percmsq}     {\mbox{cm$^{-2}$}}
\newcommand{\cii}         {\mbox{\rm [C{\small II}]}}
\newcommand{\nii}         {\mbox{\rm [N{\small II}]}}
\newcommand{\ci}         {\mbox{\rm [C{\small I}]}}
\newcommand{\wco}         {\mbox{\rm W(CO)}}
\newcommand{\fscii}       {($^2$P$_{3/2}\rightarrow^2$P$_{1/2}$)}
\newcommand{\Smol}        {\mbox{$\Sigma_{\rm mol}$}}
\newcommand{\Mmol}        {\mbox{${\rm M}_{\rm mol}$}}
\newcommand{\Ssfr}        {\mbox{$\Sigma_{\rm SFR}$}}
\newcommand{\Sgmc}        {\mbox{$\Sigma_{\rm GMC}$}}
\newcommand{\Sstar}       {\mbox{$\Sigma_*$}}
\newcommand{\mstar}       {\mbox{\rm M$_*$}}
\newcommand{\Lco}         {\mbox{$L_{\rm CO}$}}
\newcommand{\Rhmol}       {\mbox{$R^{\rm mol}_{\rm 1/2}$}}
\newcommand{\Rhst}       {\mbox{$R^*_{\rm 1/2}$}}
\newcommand{\ag}{\mbox{ \raisebox{-.4ex}{$\stackrel{\textstyle >}{\sim}$} }}
\newcommand{\al}{\mbox{ \raisebox{-.4ex}{$\stackrel{\textstyle <}{\sim}$} }}
\newcommand{\dgr}         {\mbox{$\delta_{\rm DGR}$}}
\newcommand{\dgrp}         {\mbox{$\delta_{\rm DGR}'$}}
\newcommand{\rco}{R_{\rm CO}}
\newcommand{\rht}{R_{\rm H_2}}
\newcommand{\davdg}{\Delta A_{V}}
\newcommand{\rhogas}{\mbox{$\rho_{\rm gas}$}}
\newcommand{\tsf}{\mbox{$t_{\rm SF}$}}
\newcommand{\tj}{\mbox{$t_{\rm dyn}$}}
\newcommand{\tdep}{\mbox{$\tau^{\rm mol}_{\rm dep}$}}
\newcommand{\torb}{\mbox{$\tau_{\rm orb}$}}
\newcommand{\tsh}{\mbox{$\tau_{\rm shear}$}}
\newcommand{\tshear}{\tsh}
\newcommand{\trat}{\mbox{$\tau_{\rm rat}$}}
\newcommand{\egas}{\mbox{$\Sigma_{\rm gas}$}}
\newcommand{\esfr}{\mbox{$\Sigma_{\rm SFR}$}}
\newcommand{\ehi}{\mbox{$\Sigma_{\rm HI}$}}
\newcommand{\ehtwo}{\mbox{$\Sigma_{\rm H_2}$}}
\newcommand{\emol}{\mbox{$\Sigma_{\rm mol}$}}
\newcommand{\estar}{\mbox{$\Sigma_{\rm *}$}}
\newcommand{\eightmu}{\mbox{8\,$\mu$m}}
\newcommand{\xcou}{\mbox{cm$^{-2}$~(K~km~s$^{-1}$)$^{-1}$}}
\newcommand{\Rtt}{${\cal R}_{12/13}$}
\newcommand{\f}{Fig.~}
\newcommand{\angstrom}{\mbox{\normalfont\AA}}

\date{Draft \today \hfill\fbox{\textbf{\emph{VERSION 1}}}}

\label{firstpage}

\maketitle

\begin{abstract}
We present a kpc-scale analysis of the relationship between the molecular depletion time (\tdep) and the orbital time (\torb) across the field of 39 face-on local galaxies, selected from the EDGE-CALIFA sample. We find that, on average, 5\% of the available molecular gas is converted into stars per orbital time, or $\tdep\sim 20 \ \torb$. The resolved relation shows a scatter of $\sim 0.5$ dex. The scatter is ascribable to galaxies of different morphologies that follow different \tdep$-$\torb\ relations which decrease in steepness from early- to late-types. The morphologies appear to be linked with the star formation rate surface density, the molecular depletion time, and the orbital time, but they do not correlate with the molecular gas content of the galaxies in our sample. We speculate that in our molecular gas rich, early-type galaxies, the morphological quenching (in particular the disc stabilization via shear), rather than the absence of molecular gas, is the main factor responsible for their current inefficient star formation.
\end{abstract}


%&$&$&$&$&$&$&$&$&$&$&$&$&$&$&$&$&$&$&$&$
%		Introduction
%&$&$&$&$&$&$&$&$&$&$&$&$&$&$&$&$&$&$&$&$
\section{Introduction}\label{S:intro}
Star formation is the result of an intricate interplay of dynamical, thermal, radiative, and chemical phenomena that operates on a wide range of scales (\citealt{mckee07}, \citealt{padoan2014}). Nevertheless, on a kpc scale, a simple ``recipe'' for star formation emerges. This recipe was first formalized by \cite{schmidt1959} who suggested that the star formation rate (SFR) is proportional to the square of gas volume density ($\mathrm{SFR}\propto \rho^{2}$). Afterwards, several works have aimed to empirically verify Schmidt's conjecture. The seminal studies were performed by \cite{kennicutt1989} and \cite{kennicutt1998}, who measured the relation between the surface densities of SFR ($\Sigma_{\mathrm{SFR}}$) and total gas ($\Sigma_{\mathrm{gas}}$):
\begin{equation}
\label{E:ks_rel}
\Sigma_\mathrm{SFR}\propto\Sigma_\mathrm{gas}^N,
\end{equation}
which goes by the name of the ``Kennicutt-Schmidt's relation'' (hereafter the KS relation, also called ``star formation law'') with $N=1.40\pm0.15$. This relation can also be parametrized through a single quantity called ``depletion time'' which expresses the timescale to convert the gas into stars at the current SFR:
\begin{equation}
\label{E:tdep}
\tau_\mathrm{dep} \equiv\frac{\Sigma_\mathrm{gas}}{\Sigma_\mathrm{SFR}}.
\end{equation}
The inverse of $\tau_\mathrm{dep}$ is usually called the ``star formation efficiency'' (SFE).

Resolved studies of nearby galaxies have shown that, on kpc-scales, the surface densities dominated by the molecular gas (\emol) linearly correlate with \esfr\ (e.g., \citealt{bigiel08}, \citealt{leroy2008}, \citealt{schruba11}, \citealt{leroy2013}), while the atomic gas seems irrelevant (\citealt{wong_blitz2002}, \citealt{heyer2004}, \citealt{kennicutt2007}, \citealt{schruba11}). The linearity means that the \emph{molecular depletion time} is approximately constant: \tdep$\approx2$\,Gyr (e.g., \citealt{leroy2008}, \citealt{rahman2012}, \citealt{leroy2013}). However, most of these works are carried out in local discs, but studies of the integrated star forming properties in high redshift starburst (\citealt{daddi2010}, \citealt{genzel2010}, \citealt{tacconi2010}) and quiescent early-type galaxies (\citealt{davis2014}) show that not all systems follow the same KS relation as the nearby discs.

A relation explicitly incorporating the local orbital time (hereafter \torb) is able to describe the star formation in both high-redshift starbursts and nearby discs equally well (\citealt{daddi2010}, \citealt{genzel2010}) as proposed by \cite{silk1997} and \cite{elmegreen1997}. This alternative ``star formation law'' is sometimes called the ``Silk-Elmegreen'' relation (hereafter SE relation). In particular, \cite{silk1997} defined the relation between \esfr\ and $\Sigma_\mathrm{gas}$ as follows:
\begin{equation}
\label{E:se_rel}
\Sigma_\mathrm{SFR}=\epsilon_\mathrm{orb}\frac{\Sigma_\mathrm{gas}}{\tau_\mathrm{orb}},
\end{equation}
where a fraction of gas, $\epsilon_\mathrm{orb}$ (hereafter: ``orbital efficiency''), is converted into stars during each orbital time. 

Therefore, given the SE relation, a direct proportionality between the $\tau_\mathrm{dep}$ and \torb\ is expected in the form:
\begin{equation}
\label{E:tdyn1}
\tau_\mathrm{orb} = \epsilon_\mathrm{orb} \tau_\mathrm{dep}.
\end{equation}

\cite{silk1997} described equation~\ref{E:tdyn1} as a local relation with $\tau_\mathrm{dep}=\rho_\mathrm{gas}/\rho_\mathrm{SFR}$, where $\rho_\mathrm{gas}$ is the volumetric density of total neutral gas and $\rho_\mathrm{SFR}$ is the SFR volume density. Volume densities are difficult to measure in extragalactic context, but surface densities are more easily accessed. Thus, the original \cite{silk1997} relation is generally recast in term of equation~\ref{E:se_rel}. This relation can also be applied in global terms, assuming that neutral gas and young stars have similar scale height and radial distributions.

The intuition behind this relation originally relies on two different classes of models. Following the idea of \cite{wyse1986} and \cite{wyse1989}, the multiple passages of clouds through regions of high gravitational potential (such as a spiral arm density-wave) favour the growth of clouds through collision and their subsequent collapse. Clouds on faster orbits (shorter orbital times) would encounter these gravitational depressions more frequently. As a consequence, the star formation rate and its efficiency within those objects will be enhanced with respect to clouds on slower orbits. In this case the star formation law would assume the form
\begin{equation}\label{se_rel_omegap}
\Sigma_\mathrm{SFR}\propto\Sigma_\mathrm{gas}^N(\Omega - \Omega_p), 
\end{equation}
\noindent where $\Omega=V_c/R_\mathrm{gal}$ is the disc angular speed and $\Omega_p$ is the pattern speed of the spiral arms. In the limit where $\Omega_p$ is small with respect to $\Omega$ (typically within the corotation) and $N=1$, the star formation law assumes the form of the Silk-Elmegreen relation (equation~\ref{E:se_rel}), since $\tau_\mathrm{orb}=2\pi/\Omega$.

\cite{tan2000} generalized this model to every episode of cloud compression, referring to negative shear as the main mechanism that supports the cloud coalescence in spiral arms (see also \citealt{tasker09}; \citealt{tan2010}; \citealt{suwannajak2014}). That model expressed the shear through the rotation curve shape factor $\beta=d\ln V_c/d\ln R_\mathrm{gal}$ as
\begin{equation}\label{se_rel_tan}
\Sigma_\mathrm{SFR}\propto\Sigma_\mathrm{gas}\Omega(1-0.7\beta).
\end{equation}
\noindent For a flat circular velocity curve ($\beta=0$), equation~\ref{se_rel_tan} is equivalent to the SE relation.

Alternatively, \cite{wange_silk1994} predicted that the star formation rate scales with the total amount of gas divided by the time-scale for the perturbation growth in the disc, i.e.:
\begin{equation}\label{se_rel_tgrow}
\Sigma_\mathrm{SFR}\propto\Sigma_\mathrm{gas}/\tau_\mathrm{grow}.
\end{equation}
\noindent The time-scale for the perturbation growth in a rotating disc can be expressed as \citep{tan2000}:
\begin{equation}\label{tgrow}
\tau_\mathrm{grow} \propto \sigma_\mathrm{gas}/\Sigma_\mathrm{gas}\propto Q/\kappa, 
\end{equation}

\noindent where $\sigma_\mathrm{gas}$ and $\Sigma_\mathrm{gas}$ represent the total gas  velocity dispersion and the mass surface density, respectively. The \citet{toomre64} $Q-$parameter indicates the ability of the gas to balance self-gravity with its own kinematics (parametrized via $\sigma_{\mathrm{gas}}$), together with centrifugal forces due to the disc rotation expressed through the epicyclic frequency $\kappa$. For a marginally stable disc, $Q \sim 1$ (or more generally $Q$ is constant in a self-regulating star formation scenario) so that $\kappa \propto \Omega$, $\tau_\mathrm{grow} \sim \Omega^{-1} \propto \tau_\mathrm{orb}$. Thus, Equation~\ref{se_rel_tgrow} has a similar form as Equation~\ref{E:se_rel}. Global studies (\citealt{kennicutt1998}, \citealt{daddi2010}, \citealt{genzel2010}) implicitly assume that the orbital time, measured at the outer radius of the star forming region, is equivalent to the average perturbation growth time-scale in the mid-plane of the disc.

Despite several successful observational applications (e.g., \citealt{kennicutt1998}, \citealt{boissier2003}, \citealt{daddi2010}, \citealt{genzel2010}) and theoretical derivations (e.g, \citealt{tan2000}, \citealt{krumholz_mckee2005}, \citealt{narayanan2012}), the direct proportionality between depletion time and orbital time implied by the SE relation has been often questioned. The resolved SE relation in M51 shows a significant scatter of 0.4 dex (\citealt{kennicutt2007}) and systems of different physical scales are not well described by a single SE relation (e.g. \citealt{krumholz2012}). Also, some theoretical works struggle to find such correlation (e.g., \citealt{dopita_ryder1994}, \citealt{kim2011}). 

The total gas depletion time of nearby galaxies appears proportional to the orbital time to some degree (\citealt{wong_blitz2002}), but molecular depletion time (\tdep) and \torb\ are at most weakly correlated (\citealt{leroy2008}, \citealt{wong2009}, \citealt{saintonge2011}, \citealt{leroy2013}). The correlation between \tdep\ and \torb\ seems more significant in the inner region of the galaxies, where the orbital time becomes similar to the dynamical time of the GMCs (\citealt{leroy2013}, see also \citealt{meidt2015}).

Even with these difficulties and ambiguities, the orbital time remains a way to parametrize how large-scale dynamics influence the molecular gas properties at various levels. As seen above, there remains ample theoretical motivation to explore these influences and some suggestive observational evidences. In M51, flux and integrated intensity probability distribution functions show a strong deviation from a log-normal shape in the spiral arm region (\citealt{hughes13b}). At the same time, streaming motions lengthen the molecular depletion time in the spiral arms of the galaxy (\citealt{meidt13}). The galactic disc differential rotation can set a limit to the development of expanding shells (\citealt{elmegreen2002}) and in the maximum mass that stellar clusters, GMCs, and high redshift clumps can reach (\citealt{kruijssen2014}, \citealt{reina_campos2017}). These phenomena (together with stellar feedback) might be responsible for the different slope of the GMC mass spectra in M51 environments (\citealt{colombo14a}), to create a large population of unbound clouds in the Milky Way and external galaxies (\citealt{dobbs_pringle2013}; see also \citealt{dobbs11} and references therein), and to disperse the molecular gas to large scale height (\citealt{pety2013}, \citealt{caldu_primo2013}).

Here, we expand the empirical study of the role of large-scale dynamics on molecular depletion time through a joint analysis of the Extragalactic Database for Galaxy Evolution (EDGE, \citealt{bolatto2017}) and the Calar Alto Legacy Integral Field Area (CALIFA, \citealt{sanchez2012}) surveys. This new analysis extends previous, resolved (kpc-scale) work on the local galaxy population to larger distances and a wider variety of galaxies.  Furthermore, with the high quality optical data, we are able to investigate what effects drive these relationships in the context of galaxy morphologies, stellar masses, and local properties. We focus on the parameter space defined by \tdep\ and \torb\ using a sample of 39 approximately face-on spiral galaxies (inclination $<65^{\circ}$). Given the significantly wider exploration of parameter space we can complete a rich investigation of the different factors that could influence dynamically governed star formation.

The paper is organized as follows. In Section~\ref{S:data} we summarize the two surveys CALIFA (Section~\ref{SS:califa}) and EDGE (Section~\ref{SS:edge}) and the choice of sample (Section~\ref{SS:sample}). Section~\ref{S:tdep} outlines the method we follow to obtain resolved maps of molecular depletion time from SFR and molecular gas surface densities, while in Section~\ref{S:vcirc_times} we explore the theoretical basis to calculate circular speed models and orbital time per galactocentric radius from stellar kinematics. Section~\ref{S:res_main} reports the resolved relation analyzed in the paper between \tdep\ and \torb together with its corresponding integrated version. In Section~\ref{S:torb_hubmass} the pixel-by-pixel relations are encoded via the Hubble type and the stellar mass of the galaxies to study the global parameter dependency of the main quantities. Using the same properties, we show azimuthally averaged time-scale profiles of the different morphologies in Section~\ref{S:timeprof}. Finally, we discuss and summarize our findings (in Section~\ref{S:discussion} and \ref{S:summary}, respectively). We supplement the main work with appendices that test the influence of non-detections (Appendix~\ref{A:nondetections}), local galaxy environment (Appendix~\ref{A:loc}), and the influence of atomic gas  (Appendix~\ref{A:hi}) in the analysis of the depletion time, as well as a few  additional caveats and limitations (Appendix~\ref{A:caveats}).

%&$&$&$&$&$&$&$&$&$&$&$&$&$&$&$&$&$&$&$&$
%		Data
%&$&$&$&$&$&$&$&$&$&$&$&$&$&$&$&$&$&$&$&$
\section{Data and sample}\label{S:data}
We make use of two datasets: the Calar Alto Legacy Integral Field Area (CALIFA) survey (\citealt{sanchez2012}), and the Extragalactic Database for Galaxy Evolution (EDGE) survey (\citealt{bolatto2017}). \cite{utomo2017} reports the method we used to derive SFR, molecular gas surface density, and molecular depletion time maps. The circular velocity curve models from stellar kinematics are calculated by \cite{kalinova2017b}, and the CO rotation curves are obtained by R. C. Levy et al. (in preparation). Detailed descriptions of those data products are presented in those papers. Here, we just provide a brief summary.

\subsection{The CALIFA survey}\label{SS:califa}
CALIFA is a survey that observed more than 700 SDSS galaxies in the local Universe through an integral field unit (IFU). The targets of CALIFA are chosen to be statistically representative of the galaxy population in the redshift range $0.005<z<0.03$. They cover a stellar mass range of $\log_{10}(M_{*}/[\mathrm{M_{\odot}}]) = 9.4 $-11.4 (\citealt{walcher2014}\footnote{The mass range has been originally calculated by \cite{walcher2014} using a \cite{chabrier2003} initial mass function (IMF). Throughout the paper we adopt a \cite{kroupa2001} IMF (see also \citealt{utomo2017}). By assuming the conversion factor suggested by \cite{madau_dickinson2014}, IMF$_\mathrm{Kroupa}=1.07$\,IMF$_\mathrm{Chabrier}$; therefore, the indicated mass range is consistent with our assumed IMF.}); both early and late type morphologies, including mergers, and irregular galaxies (\citealt{sanchez2016}). The data have been collected using the Postdam Multi-Aperture Spectrophotometer (PMAS) and the PMAS fiber PAcK (PPAK) IFU at the 3.5m telescope of the Calar Alto Observatory in Spain \citep{garcia_benito2015}. The CALIFA sample is diameter-selected ($45''<D_{25}<80''$), with a spatial resolution of $2.5''$ (corresponding to $\sim1$ kpc at the average redshift of the survey), and spans spectral ranges $3745 - 7300$\,\AA\ ($R\sim850$) and $3400 - 4750$\,\AA\ ($R\sim1650$). In terms of sample, CALIFA is more extended than previous, pioneering surveys (e.g. SAURON, \citealt{bacon2001} ; Atlas3D, \citealt{cappellari2011}), which imaged mainly the centre of early-type, high-mass galaxies. At the same time, the PPAK instrument ensured large spatial coverage and number of fibers: CALIFA galaxies have better linear resolution than next-generation surveys (e.g. SAMI, \citealt{croom2012}, \citealt{bryant2015}; and MaNGA, \citealt{bundy2015}, \citealt{sanchez2016}). CALIFA represents, to date, the best trade-off between sample statistics and survey design among IFU surveys.

In our analysis of CALIFA data, we use emission line maps (at \ha, \hb, [NII], [OIII] wavelengths), stellar population synthesis products, and stellar kinematic results (provided by the PIPE3D pipeline described in \citealt{sanchez2016}). These maps are regridded and smoothed to match the pixel scale and resolution of EDGE data. Low signal-to-noise ratio pixels (SNR$<2$) are blanked as well as foreground stars (see \citealt{utomo2017} for details).

\subsection{The EDGE survey}\label{SS:edge}
EDGE is the $^{12}$CO $J=1 - 0$ and $^{13}$CO $J=1 - 0$ follow-up survey of 177 infrared-bright CALIFA galaxies. Among those, 126 targets have been observed with both D+E configurations of CARMA, yielding a typical synthesized beam of $\sim4.5''$. Given the average distance of EDGE targets, this resolution corresponds to approximately $1.5$\,kpc. The D+E cubes used in this analysis have a channel width of 10\,\kms. The RMS noise in EDGE cubes ranges between $40 - 65$\,mK with an average of $\sim50$\,mK, giving a $3\sigma_\mathrm{RMS}$ sensitivity of $\Sigma_\mathrm{mol}\sim11$\,M$_\odot$\,pc$^{-2}$ (before inclination corrections). EDGE is by far the most extended interferometric $^{12}$CO(1-0) survey in the local Universe. It spans broader ranges of color, luminosity, stellar masses, and morphological types than the previous kpc-scale surveys which focused on nearby, blue, star forming galaxies (e.g. BIMA SONG, \citealt{regan01}; \citealt{helfer03}; HERACLES, \citealt{leroy2009}; CARMA STING, \citealt{rahman2011}, \citealt{rahman2012}) or the centers of early type galaxies (\citealt{alatalo2013}).  Full information about the EDGE survey design and data reduction appeared in \citet{bolatto2017}.

In our studies, we use moment maps from D+E data cubes, masked in order to capture CO emission. The signal-to-noise ratio in pixels within the mask is mostly SNR$>2$. Those masks have been constructed using the {\sc IDL} method developed by \citet{wong2013}\footnote{{https://github.com/tonywong94/idl\_mommaps}}. To minimize oversampling, the original EDGE images have been regridded so that each independent resolution element corresponds to 4 pixels (\citealt{utomo2017}). This sampling is used throughout this paper. 

\begin{figure*}
\centering
{\includegraphics[width=\textwidth]{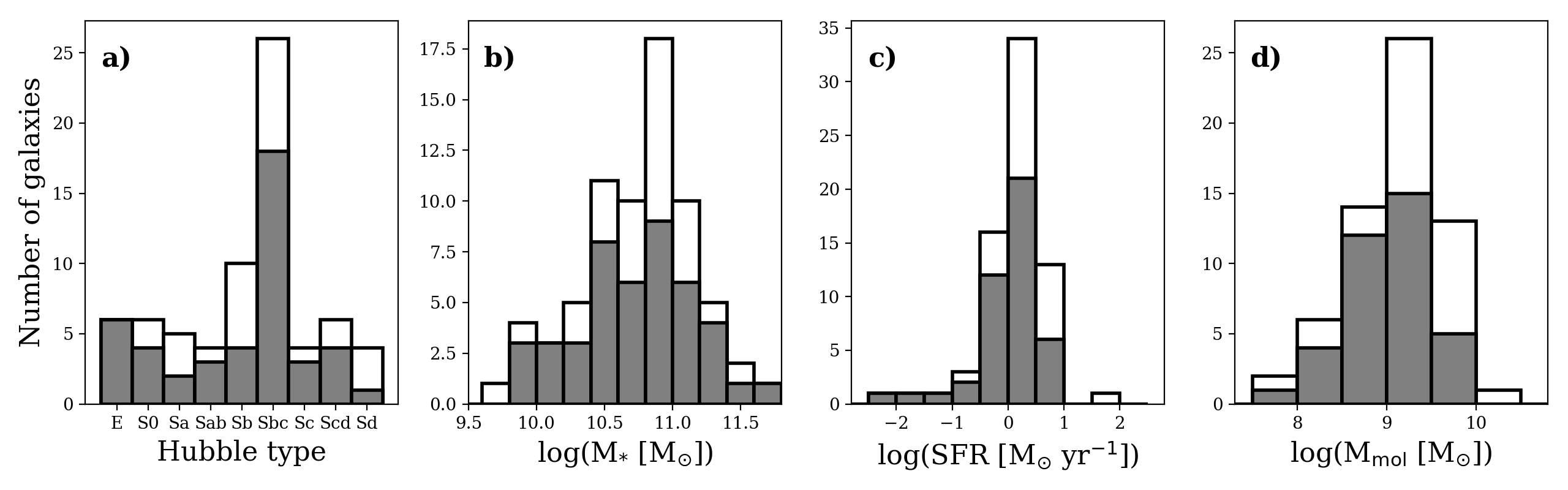}}
\caption{The histograms of Hubble types (panel $a$; \citealt{walcher2014}), integrated stellar masses (panel $b$; \citealt{sanchez2016}), SFR (panel $c$; \citealt{bolatto2017}), and molecular gas masses (panel $d$; \citealt{bolatto2017}) of the full EDGE D+E sample (white) and the sample in this paper (gray). Both samples are restricted to galaxies with inclination below $65^{\circ}$. This figure shows that the sample that is  selected for this study is still representative of the full EDGE sample of \citet{bolatto2017}.}
\label{F:sample}
\end{figure*}

\subsection{Sample selection}\label{SS:sample}
The sample considered in this paper consists of 83 objects, which is the overlap between 126 galaxies from the full EDGE D+E sample and the 238 CALIFA galaxies that have the circular velocity curve modeled by \cite{kalinova2017b}. In most of the analysis, we only use galaxies with inclination below 65$^{\circ}$. We explore the effect of inclinations in highly inclined galaxies in Appendix~\ref{A:caveats}. There are 71 EDGE galaxies with inclination below this 65$^{\circ}$ limit, which, when restricted to the samples with dynamical models and that encompass at least one line-of-sight with SNR>2 (for both \esfr\ and \emol) results in 39 objects. Fig.~\ref{F:sample} shows the comparison between the parent EDGE sample (after inclination cut) and the sample used in this paper. Although there is a significant reduction in the number of galaxies, our sample is still a representation of the EDGE sample (after inclination cut) in terms of the Hubble type, stellar masses, star formation rates, and molecular gas masses. However, very late-type galaxies (Sd) are underrepresented in our sample. Galaxies with log(SFR/$\msunperyr$) $>1$ and log($M_\mathrm{mol}/\msol$) $>10$ are also underrepresented with respect to the EDGE sample (after inclination cut).

\begin{table*}
\scalebox{0.85}{
\begin{tabular}{cccccccccccc}
\hline
\hline
Galaxy & Hubble type & $\Sigma_\mathrm{SFR}$ & $\Sigma_\mathrm{mol}$ & $\tau_\mathrm{dep}^\mathrm{mol}$ & $\tau_\mathrm{orb}$ & $A$ & $R_\mathrm{eff}$ & $i$ & $D$ & $N_\mathrm{los}^\mathrm{det}$ & $N_\mathrm{los}^\mathrm{non-det}$ \\
& & [$10^{-4}$\,M$_\mathrm{\odot}$\,kpc$^{-2}$\,yr$^{-1}$] & [M$_\mathrm{\odot}$\,pc$^{-2}$] & [$10^{9}$\,yr] & [$10^{8}$\,yr] & [$10^{-8}$\,yr$^{-1}$] & ["] & [$^{\circ}$] & [Mpc] & & \\  
(1) & (2) & (3) & (4) & (5) & (6) & (7) & (8) & (9) & (10) & (11) & (12)\\  
\hline
NGC6021 & E5 & $0.65\pm0.01$ & $19.56\pm0.69$ & $302.53\pm7.28$ & $2.41\pm0.07$ & $1.23\pm0.04$ & 11.18 & 43 & 69 & 2 & 14 \\
NGC5485 & E5 & $1.4\pm0.1$ & $13.31\pm1.56$ & $109.11\pm20.46$ & $0.53\pm0.08$ & $1.89\pm0.21$ & 15.79 & 47 & 27 & 7 & 32 \\
NGC5784 & S0 & $136.47\pm64.47$ & $24.83\pm8.31$ & $2.52\pm1.13$ & $0.69\pm0.22$ & $4.28\pm1.06$ & 12.04 & 45 & 79 & 73 & 148 \\
... & ... & ... & ... & ... & ... & ... & ... & ... & ... & ... \\
... & ... & ... & ... & ... & ... & ... & ... & ... & ... & ... \\
... & ... & ... & ... & ... & ... & ... & ... & ... & ... & ... \\
NGC3381 & Sd & $165.1\pm81.93$ & $21.98\pm4.68$ & $1.47\pm0.76$ & $1.53\pm0.69$ & $0.35\pm0.08$ & 21.42 & 31 & 23 & 59 & 209 \\
\hline
\hline
\end{tabular}}
\caption{The sample of nearby EDGE-CALIFA galaxies considered in this work: (1), CALIFA name of the galaxies; (2), Hubble type defined by eye from members of the CALIFA team as described in Walcher et al. 2014. In the paper we group different elliptical types in the same category ``E'', S0 and S0a galaxies are considered as ``S0'' type, while Sdm galaxies are grouped together to ``Sd'' types; (3), median and median absolute deviation of the detected lines-of-sight from the SFR surface density map of a given galaxy; (4), median and median absolute deviation of the detected lines-of-sight from the molecular gas mass surface density map of a given galaxy; (5), median and median absolute deviation of the detected lines-of-sight from the molecular depletion time map of a given galaxy; (6) median and median absolute deviation of the orbital time where both \esfr\ and \emol\ are detected; (7), median and median absoluted deviation of the local shear rate (Oort's A) where both \esfr\ and \emol\ are detected; (8), effective radius measured by Walcher et al. 2014; (9), inclination from CO kinematics (R. C. Levy et al. in prep.); (10), distance from HyperLEDA; (11), number of detected \tdep\ lines-of-sight (both \esfr\ and \emol\ detected with SNR>2); (12), number of non-detected \tdep\ lines-of-sight (either \esfr\ or \emol\ have SNR<2).}
\label{T:sample}
\end{table*}

%&$&$&$&$&$&$&$&$&$&$&$&$&$&$&$&$&$&$&$&$
%     Molecular depletion time maps
%&$&$&$&$&$&$&$&$&$&$&$&$&$&$&$&$&$&$&$&$
\subsection{Molecular depletion time maps}\label{S:tdep}
The maps of molecular depletion time have been generated by \cite{utomo2017} and they are fully described there. Here, we give a short summary of their derivation method. The molecular depletion time is calculated by:
\begin{equation}
\tdep = \frac{\emol}{\esfr}.
\end{equation}
\noindent The deprojected maps of molecular gas surface density (\emol) are generated by following \cite{leroy2008}:
\begin{equation}\label{E:Sh2}
\frac{\emol}{\msunperpcsq} = \aco\,\cos i\,\frac{I_{\mathrm{CO}}}{\mathrm{K}\,\kms},
\end{equation}
where we assume the Galactic CO-to-H$_{2}$ conversion factor of $\aco=4.4$ to convert between $^{12}$CO ($J=1-0$) integrated intensity ($I_{CO}$) to \emol; $\cos i$ accounts for the deprojected area due to the inclination ($i$) of galaxies.

The maps of star formation rate surface density (\esfr) are obtained from:
\begin{equation}\label{E:SSFR}
\esfr(\msunperpcsqyr) = C\times\frac{4\pi d^2 F(H_{\alpha})\cos i}{S_\mathrm{pix}}\times10^{0.4 \ A_{H_{\alpha}}}.
\end{equation}
The \ha\ flux maps ($F(H_{\alpha})$) are converted to \ha\ luminosity by considering the distance $d$ of the galaxies in pc, which are drawn from the HyperLEDA catalog (\citealt{makarov2014}). The nebular extinction of \ha\ ($A_{H_{\alpha}}$) is calculated applying the Balmer decrement method (e.g., \citealt{catalan_torrecilla2015}, equation 1), which compares the observed and theoretically expected ratios between \ha\ and \hb\ fluxes. To convert between extinction-corrected \ha\ luminosities and SFR, we use the calibration factor $C=5.3\times10^{-43}$\,(\msunperyr\,erg$^{-1}$\,s) from \cite{calzetti2007}. Finally, SFR surface density maps are generated by dividing SFR by the physical pixel area of the regridded and deprojected \ha\ flux maps, $S_\mathrm{pix}$, expressed in kpc. Additionally, we blank the AGN-like emission pixels that lie above the \cite{kewley_dopita2002} relation in the BPT diagram (constructed by the [NII]/\ha\ and [OIII]/\hb\ line ratio maps) as well as pixels with \ha\ equivalent width $<6$\angstrom, because this \ha\ emission is caused by stars older than 500 Myrs, which are not associated with star formation (\citealt{sanchez2014}).

Pixels within \esfr\ and \emol\ maps with SNR $>2$ are considered as detections in our analysis. Following \cite{utomo2017}, we will also study what we consider as upper and lower limits of the depletion time. Upper \tdep\ limits are measured by including non-detections in \emol\ as 2$\sigma_\mathrm{rms}$ values. The uncertainty level within the masked cube (1$\sigma_\mathrm{rms}$) is calculated as the standard deviation of the noise along the velocity axis. The lower limits are given by the non-detections of the SFR surface density. We considered non-detections in \esfr\ as values below 2$\sigma_\mathrm{rms}$, where the noise level (1$\sigma_\mathrm{rms}$) is determined by the median absolute deviation of the AGN-masked CALIFA H$_{\alpha}$ maps.

%&$&$&$&$&$&$&$&$&$&$&$&$&$&$&$&$&$&$&$&$
%     Circular velocity models
%&$&$&$&$&$&$&$&$&$&$&$&$&$&$&$&$&$&$&$&$
\subsection{Dynamical models, circular velocity curves, and orbital times}\label{S:vcirc_times}
We calculate the orbital times from the circular velocity curves (CVCs) inferred from stellar kinematics and SDSS r-band surface brightness \citep{kalinova2017b}, rather than from the CO rotation curve. This choice maximizes the dynamic ranges of stellar masses and Hubble types in our sample, because the rotation curves inferred from CO are mainly available for intermediate morphological types (R. C. Levy et al. in preparation). Since stars are present in all galaxy types, we can obtain orbital times for EDGE targets that are barely resolved in CO emission without being constrained by the accuracy of their CO rotation curves. A similar approach has been used by \cite{davis2014} to calculate rotation curves for a sample of fast rotating early-type galaxies.

The CVCs were derived using the axis-symmetric case of Jeans anisotropic multi-Gaussian expansion dynamical model (JAM\footnote{http://www-astro.physics.ox.ac.uk/~mxc/software/\#jam}; \citealt{cappellari2008}). In the JAM approach, two basic assumptions are made for the stellar population: a constant velocity anisotropy ($\beta_\mathrm{z}=1-\sigma^2_\mathrm{z}/\sigma^2_\mathrm{R}$) and a constant dynamical mass-to-light ratio ($\Upsilon_{\mathrm{dyn}}$).
The parameters $\beta_\mathrm{z}$ and $\Upsilon_{\mathrm{dyn}}$ have been defined after fitting the observed second-order velocity moment $V_{\mathrm{rms}}=\sqrt{V^2+\sigma^2}$ calculated from the stellar kinematics of the galaxies (\citealt{falcon-barroso2016}).
The best-fit model of the observed $V_{\mathrm{rms}}$, the corresponding fitting parameters
($\beta_\mathrm{z}$ and $\Upsilon_{\mathrm{dyn}}$), and their uncertainties are obtained by applying the Markov Chain Monte Carlo (MCMC) method as described in \citet{kalinova2017a}.

The circular velocity curve is derived by applying Poisson's equation to the best fit of gravitational potential $\Phi(R,z)$.
$\Phi(R,z)$ is generated via the multi-Gaussian expansion method (MGE; \citealt{monnet1992}; \citealt{emsellem1994}), where the observed surface brightness of the galaxies is parametrized as a sum of $N$ Gaussian components, which represent the photometry of the galaxies in detail, as follows:
\begin{equation}
  \label{E:mgeSB}
  I(x',y') = \sum_{j=0}^{N} I_{0,j} \exp\left\{ -\frac{1}{2{{\bf \xi}'_j}^2} \left[ x'^2 + \frac{y'^2}{{q'_j}^2} \right] \right\},
\end{equation}
where $I_{0,j}$ is the central surface brightness, ${\bf \xi}'_j$ is the dispersion along the major $x'$-axis
and $q'_j$ describes the flattening of the ellipses. 
The intrinsic dispersion $\xi_j$ and flattening $q_j$ are related to their observed (plane-of-sky) equivalents to:
\begin{equation}
  \label{E:qintr}
  	\xi_j = \xi'_j
  	\quad \mathrm{and} \quad
  	{q'_j}^2 = \cos^2 i + q_j^2 \sin^2 i.
\end{equation}

\cite{kalinova2017b} obtained the MGE models using the software implementation of \cite{cappellari2002}. The code was applied to the $r$-band photometric images from the Sloan Digital Sky Survey (SDSS; \citealt{york2000}) using Data Release $12$ (DR12; \citealt{alam2015}).

Finally, the circular velocity from the JAM model is derived from (e.g., Section 3.2. of \citealt{kalinova2017a}):
\begin{multline}
 \label{E:mgevcirc}
 V_c^2(R_\mathrm{gal})\equiv V_{c,\mathrm{JAM}}^2(R_\mathrm{gal}) = \sum_{j=0}^{N} 
 \frac{2 G L_j \Upsilon_j}{\sqrt{2\pi}\xi_j} \frac{R_\mathrm{gal}^2}{\xi_j^2} 
  \times \\
\int_0^1 \exp\left\{ -\frac{u^2 R_\mathrm{gal}^2}{2\xi_j^2} \right\}
\frac{u^2\,\mathrm{d}u}{\sqrt{1-(1-q_j^2)u^2}},
\end{multline}
where $L_j \equiv 2 \, \pi \xi_j^2 q'_j I_{0,j}$ and $\Upsilon_j$ are the total luminosity and the mass-to-light ratio of the $j$th Gaussian. 
Due to the assumption of a constant mass-to-light ratio,  $\Upsilon_j$ is taken to be the same for all Gaussians in the dynamical model, 
i.e., $\Upsilon_j=\Upsilon_\mathrm{dyn}$, $\forall j$ (see Section 3.2 of \citealt{kalinova2017a}).  

Given the circular velocity curve, we can compute the orbital time at each galactocentric radius $R_{\mathrm{gal}}$ as:
\begin{equation}\label{E:tdyn}
\torb (\yr) = \frac{2\pi R_{\mathrm{gal}}}{V_c(R_{\mathrm{gal}})}.
\end{equation}
The orbital time is a proxy for the depth of the global potential well of the galaxies within $R_{\mathrm{gal}}$. We ensure that all galaxies in our sample are fast rotators by performing the test described in \cite{emsellem2011} equation 3, where galaxy ellipticity and angular momentum are calculated by \cite{kalinova2017b} (see also Falc\'on-Barroso et al. in preparation). Moreover, JAM circular speeds are largely in agreement with the CO rotation curves. Comparing JAM models to CO curves only reveals small differences (typically $\sim10$ \kms, G. Leung et al. submitted), despite the assumption of a constant $\Upsilon_\mathrm{dyn}$ that has been shown to be not completely appropriate for galactic discs (\citealt{de_denus_Baillargeon2013}). \torb\ estimates reflect the orbital times within the molecular gas in the mid-plane. Nevertheless, we discuss in Appendix~\ref{A:caveats} possible biases introduced by the JAM modeling in the \tdep-\torb\ relation.

\begin{figure*}
\centering
{\includegraphics[width=0.85\textwidth]{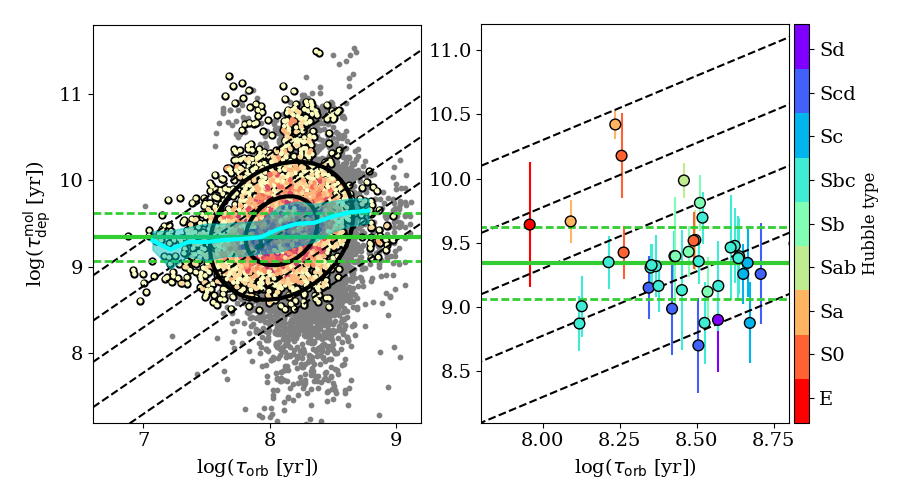}}
\caption{\emph{Left}: The resolved relationship between \tdep\ and \torb\ in our EDGE-CALIFA sample. This figure shows that the detected lines-of-sights cluster around the 5\% orbital efficiency with a large scatter of 0.5 dex. Coloured circles in the diagrams refer to the detections only, with darker colours represent higher number densities of pixels. The upper and lower limits of \tdep\ are indicated as gray points. Black ellipses mark the $1\sigma$ (inner) and $2\sigma$ (outer) confidence interval of the data, derived from the Principal Component Analysis (see text). The cyan line and its band indicate the median and interquartile range of \tdep\ within bins of 0.2 dex in \torb. \emph{Right}: The integrated measurements of the molecular depletion times (calculated within 2\,$R_\mathrm{eff}$) and the orbital times (measured at 2\,$R_\mathrm{eff}$). The measurements of \tdep\ and \torb\ in Sb-Sbc galaxies (which dominate our sample) are moderately correlated (Spearman rank $\sim0.7$) and show a median orbital efficiency of 10\%, while other Hubble type galaxies largely deviate from this value. Considering measurements within different effective radii does not change the general trend with the Hubble types that we show in this figure. In both panels, parallel dashed black lines represent loci where a certain fraction of molecular gas is consumed by star formation at each orbit (orbital efficiency, $\epsilon_\mathrm{orb}$): 0.5\%, 1.7\%, 5\%, 17\%, and 50\% (from top to bottom). As a comparison, the horizontal solid green line marks the average depletion time of 2.2 Gyr as measured in nearby spiral galaxies by \citealt{leroy2013}, together with their scatter of $\pm0.28$ dex (dashed green lines).}
\label{F:tdep_torb}
\end{figure*}

%&$&$&$&$&$&$&$&$&$&$&$&$&$&$&$&$&$&$&$&$
%           Results
%&$&$&$&$&$&$&$&$&$&$&$&$&$&$&$&$&$&$&$&$
\section{A relation between the resolved molecular depletion and orbital times from local galaxies?}\label{S:res_main}
Identifying the connections between the molecular depletion time and other timescales in the galaxies can give insight into the physics that regulates the star formation. The orbital time is the longest of the relevant dynamical times in the galactic discs (see, e.g. \citealt{semenov2017} and references therein), therefore it is the closest in magnitude to typical molecular depletion time values measured on kpc-scale. Here, we explore the connections between molecular depletion time and orbital time using resolved measurements from EDGE and CALIFA data. Fig.~\ref{F:tdep_torb} (left) shows the result of the analysis as a bi-dimensional histogram. The data are largely scattered around the following values:
\begin{itemize}
\item \torb=(3.2$^{+2.0}_{-1.2}$)$\times10^8$\,yr,
\item \tdep=(2.8$^{+2.3}_{-1.2}$)$\times10^9$\,yr;
\end{itemize} 
where the characteristic values are the median of the respective distributions, and the scatter is given by the interquartile range, spanning the 25$^\mathrm{th}$ to 75$^\mathrm{th}$ percentiles of the distributions. Beside the difference in order of magnitude, the depletion and orbital times show similar dynamic ranges. Molecular depletion time values are largely in agreement with the results of \cite{leroy2013} which measure a \tdep=(2.2$^{+2.0}_{-1.0}$)$\times10^9$\,yr. Their sample, however, does not include early-type galaxies, which can shift the \tdep\ median to higher values. In the panel, the two concentric $1\sigma$ and $2\sigma$ confidence ellipses approximate the regions of the diagram that contain $\sim68\%$ and $\sim95\%$ of the data, respectively; oriented in the direction of maximal variance of the data points. Those ellipses are obtained by performing a Principal Component Analysis (PCA) of the data. The PCA technique (\citealt{pearson1901}) constructs the covariance matrix of the data and performs a spectral embedding of the matrix in order to describe the data through their larger variance direction. The confidence ellipsoids are oriented by the main eigenvector of the matrix (i.e. the eigenvector with the largest eigenvalue) which indicates the direction of maximum extension of the data in the plane. The orientation of the main eigenvector defines, in essence, the \emph{slope} of the relation under analysis. The major and minor axes of the ellipsoid are calculated as $\sigma_\mathrm{maj, min} = 2\sqrt{\lambda_\mathrm{maj, min}}$, where $\lambda_\mathrm{maj, min}$ indicate the largest and the smallest eigenvectors of the matrix, respectively. In this case, $\sigma_\mathrm{min}$ represents the \emph{scatter} of the relation. The angle between the ellipses and the x-axis is $\sim70^{\circ}$ (equivalent to a slope $\sim3.4$), indicating that the relation between the two \tdep\ and \torb\ is much steeper than linear. Also, the two quantities are not strongly correlated given that the ratio between the $1\sigma$ ellipse major and minor axes is $\sim1.5$. Despite the scatter, an orbital efficiency of $\epsilon_\mathrm{orb} = 5\%$ describes quite well the trend in the data and follows the median of the data within the $1\sigma$ ellipse. The $1\sigma$ ellipse is bounded by the $\epsilon_\mathrm{orb}=1.7\%$ and $\epsilon_\mathrm{orb}=17\%$ lines. Globally, therefore, the resolved measurements of depletion time appear to follow equation~\ref{E:tdyn1} (restricted to the molecular gas mass surface density) with $\epsilon_\mathrm{orb}=5\%$ and a scatter of about $\sim0.5$\,dex; in other words $5\%$ of the available molecular gas is converted into stars at each orbit in our galaxies. The data are asymmetrically scattered around the outer confidence ellipse toward high values of depletion time, probably due to the shortage of late type galaxies in our sample. Our orbital efficiency value is comparable to other results from local galaxies in the literature which found $\epsilon_\mathrm{orb}\sim6-7\%$ (\citealt{kennicutt1998}, \citealt{wong_blitz2002} , \citealt{leroy2008}). Nevertheless, a clear correlation between molecular depletion time and orbital time is rarely observed (e.g., \citealt{wong2009}, \citealt{leroy2008}) especially in the molecular-dominated regime (\citealt{leroy2008}). In order to check the impact of \emol\ and \esfr\ non-detections, we include upper and lower limits of the molecular depletion time in the plots (gray points). By adding those values, the relation between \tdep\ and \torb\ becomes highly questionable, since many low molecular depletion time points are added. Those values are mostly due to \emol\ non-detections (see Appendix~\ref{A:caveats}).

Originally, the relation between depletion time and orbital time has been tested indirectly by \cite{kennicutt1998} using total gas mass surface densities and integrated measurements. In his study, the assumed orbital time was measured at the outer edge of the star-forming region and it was used as dynamical time in his version of the Schmidt's law. The author concluded that the Schmidt's relation, modified to account for the orbital time, provides a feasible star formation law. \cite{kennicutt1998} measured an orbital efficiency of 10\% in a sample of normal spiral plus starburst galaxies. Following this seminal work, we use the orbital time measured at 2\,$R_\mathrm{eff}$, together with  molecular depletion times integrated within the same radius (Fig.~\ref{F:tdep_torb}, right). Additionally, we color-encode the data points by the Hubble type of their galaxies. The morphology of our galaxies has been defined by-eye by members of the CALIFA teams as described in \cite{walcher2014}. Sbc galaxies which, in number, dominate our sample (we have 18 Sbc on a total of 39 objects) appears to cluster around $\epsilon_\mathrm{orb}\sim10\%$. Moreover the data points belonging to these galaxies appear moderate correlated showing a Spearman rank $\sim0.7$. Including the Sb galaxies (which follow the same orbital efficiency of 10\%) the Spearman rank decreases to 0.5. Nevertheless, other type galaxies largely deviate from this value. In particular the early-types (e.g, E, S0, and Sa types) show very long depletion times, away from the main relation. Across the Hubble sequence global measurements of \tdep\ and \torb\ appear actually anti-correlated. A similar anti-correlation has been noticed by \cite{leroy2013} (see their Figure 7) for a different sample of nearby discs, which, as in our case, include only molecular gas. 

For comparison with the analysis of \cite{kennicutt1998} and \cite{leroy2008}, we simulate the effects of including atomic gas in our analysis. The results of the test, reported in Appendix~\ref{A:hi}, suggest that considering the atomic gas might not significantly alter the appearance of both pixel-by-pixel and integrated depletion time - orbital time relationships in our molecular-rich galaxies.

\begin{figure*}
\centering
{\includegraphics[width=1\textwidth]{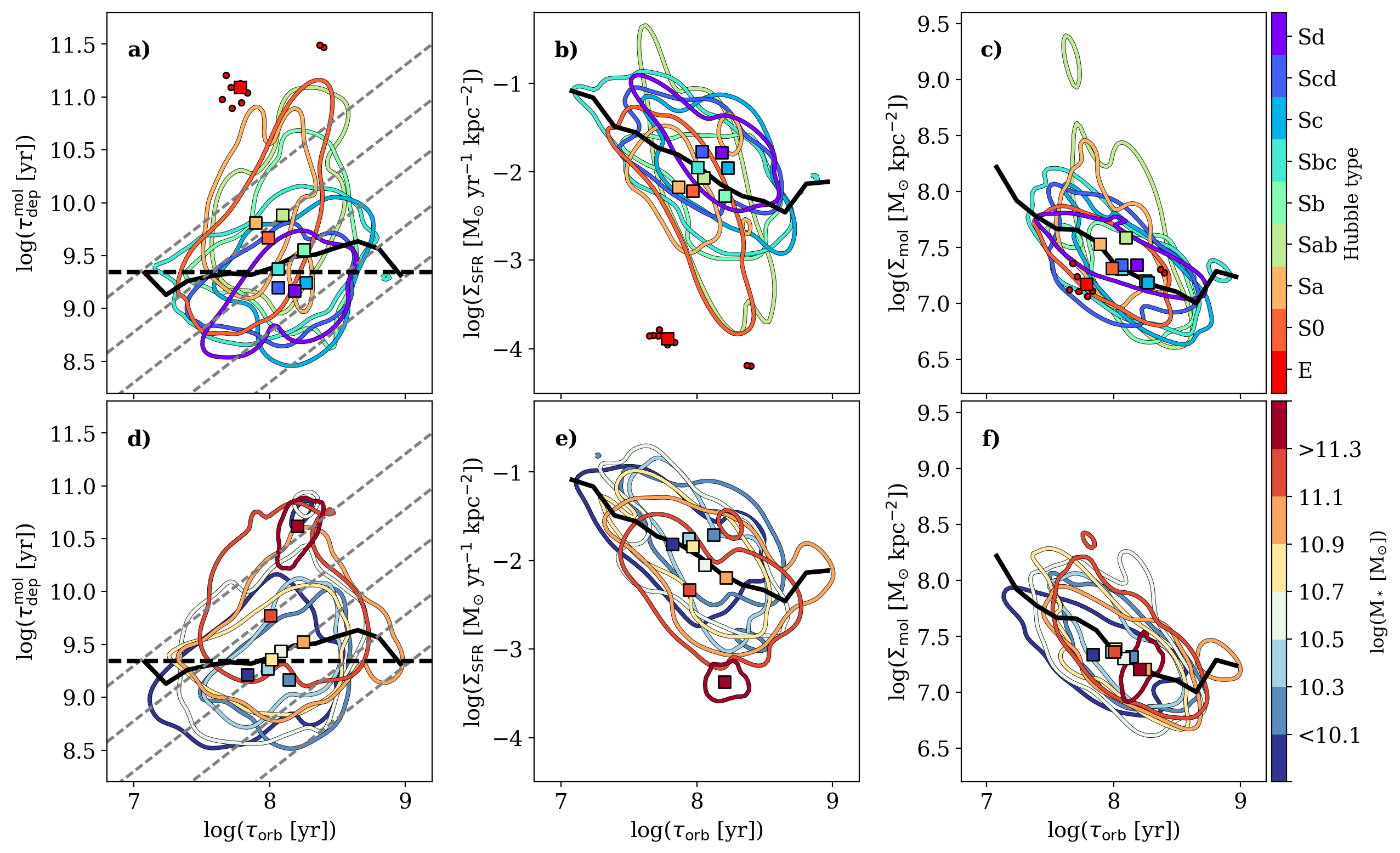}}
\caption{Pixel-by-pixel relations between \tdep\ (first column), \esfr\ (second column), and \emol\ (third column) with respect to \torb, where both \esfr\ and \emol\ are detected with line-of-sights SNR$>2$. The data are color-encoded by Hubble type (first row) and integrated stellar mass (second row). The colored contours indicate the Kernel Density Estimation (KDE)-smoothed surface that contains 95\% of the points in a given category. Elliptical galaxies, for which we possess only few line-of-sights, are plotted with red circles, instead of a contour. Colored squares indicate the positions of the quantity medians for each category. The black solid lines indicate the medians of the related quantities within 0.2 dex bins of orbital times. The black horizontal dashed lines in panels $a$ and $d$ mark the nearby star forming galaxy depletion time of 2.2 Gyr \citep{leroy2013}, while the gray diagonal lines are the constant orbital efficiencies (from left to right) of 0.5\%, 1.7\%, 5\%, 17\%, and 50\%. This figure shows that the \tdep$-$\torb\ and \esfr$-$\torb\ relations (panels $a$ and $b$) are segregated by the Hubble types, in the sense that the slopes of the relation are shallower from early to late-type galaxies. However, these trends can not be simply attributed to the different stellar masses of galaxies (panels $d$ and $e$) nor the molecular gas surface density (panels $c$ and $f$).}
\label{F:hubmassplot}
\end{figure*}

\section{Resolved orbital time relationships across morphologies and masses}\label{S:torb_hubmass}

In the previous section, we show that the resolved \tdep\ and \torb\ measurements from our sample of EDGE-CALIFA galaxies cluster around an orbital efficiency of 5\%, albeit with a scatter of 0.5 dex. In addition, we showed that the integrated molecular depletion times in early-type galaxies deviate from the \tdep$-$\torb\ relation of the later type galaxies. In this regard, we want to test if the scatter in the resolved \tdep-\torb\ relationship is attributable to the Hubble type by color-encoding the detected line-of-sight by their morphologies. The result of this analysis is shown in Fig.~\ref{F:hubmassplot}, where we also consider \esfr\ and \emol\ versus \torb . When the galaxies are segregated through the Hubble types, the medians of the depletion and orbital times reflect the behavior of the integrated measurements, i.e. the kpc-size regions in the early-type galaxies have longer depletion times, but shorter orbital times with respect to the late-types (Fig.~\ref{F:hubmassplot}$a$). In particular, the medians for Sb-Sbc galaxies lie exactly on the $\epsilon_\mathrm{orb}=5\%$ locus. Even though the averaged \esfr\ and \torb\ correlate with morphologies (Fig.~\ref{F:hubmassplot}$b$), it is not true for \emol, where the distribution of median values appear to be packed at the same location (Fig.~\ref{F:hubmassplot}$c$). Those trends are more prominent if non-detections are included (see Appendix~\ref{A:nondetections}).

Additionally, different Hubble types appear to follow different \tdep - \torb\ relations relative to from the global trend (Fig.~\ref{F:hubmassplot}$a$). For earlier-types (e.g, S0, Sa, Sab) the correlations are steep. This steepness decreases in the late-types (e.g, Sc, Scd, and Sd). For Sb galaxies, the molecular depletion time appears to be uncorrelated with \torb. The latter result is consistent with comparable, kpc-scale, resolved studies (e.g., \citealt{leroy2008}, \citealt{wong2009}, \citealt{leroy2013}; which include late spirals only) that do not identify a significant trend between the two quantities. The difference in the steepness of the \tdep - \torb\ relation with the Hubble types seems to be driven by the behavior of \esfr\ with the orbital time. Indeed, the \esfr - \torb\ relationship flattens from early- to late-types, while \emol\ and \torb\ are anti-correlated, but, modulo outliers, no significant variations with the Hubble types are observed. Therefore, regarding our sample, the different behaviors of molecular depletion time with respect to orbital time and galaxy morphology are mostly driven by the correlations of SFR surface densities with these two parameters (\tdep\ and \torb).

Similarly, data points corresponding to different Hubble types are well localized in particular regions of the diagrams. Regions above the nearby galaxy depletion time (\tdep$\sim 2$~Gyr) and above the sample running average are mostly populated by early-type galaxies (e.g, E, S0, and Sa). Elliptical galaxies, for which we possess only a few measurements, show very large values of depletion time (\tdep $>10^{11}$\,yr). The early-type galaxies seem to have somewhat shorter orbital times than others. Sab, Sb, and Sbc span larger regions of the \tdep$-$\torb\ diagram, where the depletion time decreases from Sab to Sbc types, reaching value of an order-of-magnitude below the sample average and the nearby galaxy value. At the same time, late-type galaxies show longer orbital time compared to the early types (\torb $>10^{8.5}$\,yr) that slowly increases from Sab to Sbc. Nevertheless, those galaxy morphologies cover the largest area of the \tdep-\torb\ diagram. Most of the late-type galaxy molecular depletion times (e.g., Scd and Sd) are, instead, below the average. For these galaxies orbital times span a large range ($10^{6.5}<$ \torb $<10^{9.2}$\,yr).

If we consider the average orbital efficiency of the full sample (i.e. $\epsilon_\mathrm{orb}=5\%$), the scatter of the detected lines-of-sight across the Hubble sequence decreases: $\sim2$\,dex for E galaxies, $\sim0.7$\,dex for S0, Sa, and Sab galaxies; and $\sim0.3-0.4$\,dex for galaxies between Sb to Sd types.

\cite{gonzalez_delgado2015} clearly showed that the stellar mass of galaxies decreases along the Hubble sequence (see their Figure 2). To understand whether the correlations with morphology are merely a reflection of the stellar mass behavior, in the second row of Fig.~\ref{F:hubmassplot} we group the lines-of-sight within bins of 0.2 $M_*$ dex. Stellar masses are  obtained from the product of stellar population synthesis of \cite{sanchez2016} (after converting from Salpeter to Kroupa IMF). Fig.~\ref{F:hubmassplot}$d$ shows that the molecular depletion time increases with increasing stellar mass. High stellar masses ($M_*>10^{11}$\,M$_{\odot}$) correspond to \tdep$\gtrsim 10^9$\,yrs, while low mass galaxy data span the full parameter space: galaxies with stellar masses $<10^9$\,M$_{\odot}$ can also have $10^{8.5}<$\tdep$<10^{11}$\,yrs. This result resembles what was found by Bolatto et al. (2017; see their Fig. 18), where galaxies with stellar mass above (below) $M_*=10^7$\,M$_\odot$ appear to dominate the depletion time values above (below) the full sample \tdep\ median (see also Fig.~\ref{F:hubmassplot}$d$). 

Nevertheless, the trend of the distribution of medians in the various diagrams is less clear when segregated through stellar mass bins. The depletion and orbital times are weakly correlated across the stellar masses (Fig.~\ref{F:hubmassplot}{\it d}), as well as \esfr\ and \torb\ which seems to be anti-correlated in the same representation (Fig.~\ref{F:hubmassplot}$e$). As for the galaxy morphologies, we do not distinguish any clear trends when \emol\ is considered (Fig.~\ref{F:hubmassplot}{\it f}).

Data points within different $M_*$ bins do not follow different \tdep-\torb\ relationships as for the Hubble type. The behavior of the depletion time with the stellar mass seems to be driven mostly by the SFR per unit area that clearly decreases with increasing stellar mass. Again, data points corresponding to low mass galaxies ($M_*<10^9$\,M$_{\odot}$) span all \esfr\ allowed by our data. The \emol-\torb\ relationship and \emol\ itself do not look significantly influenced by stellar mass (Fig.~\ref{F:hubmassplot}$f$).

\begin{figure}
\centering
{\includegraphics[width=0.5\textwidth]{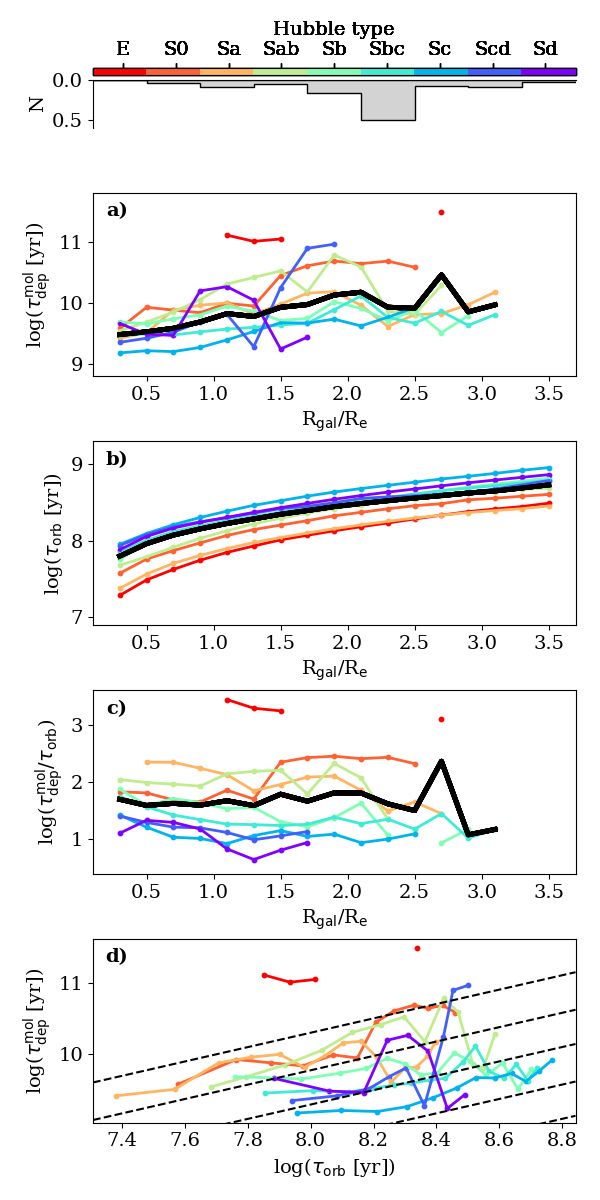}}
\caption{The azimuthal averages of \tdep\ (panel $a$), \torb\ (panel $b$), \tdep/\torb\ (panel $c$) as a function of effective radii. The azimuthal average is defined as the median of the quantities in a particular morphological type within the radial bins of 0.2\,\rgal/\reff. Panel $d$ shows the azimuthal average of \tdep\ versus \torb\ where the dashed lines indicate constant orbital efficiencies of 0.5\%, 1.7\%, 5\%, 17\%, and 50\%. The solid black lines in panels $a$, $b$, and $c$ are the average (median) profiles of the sample. The histogram on the top row of the figure indicates the relative fractions of pixels within a particular Hubble category. This figure shows that the molecular depletion time profiles from early-type (late-type) galaxies tend to be located above (below) the sample average, while the opposite behavior is observed in the orbital time azimuthal averages.}
\label{F:profplot}
\end{figure}

\section{Time-scale profiles}\label{S:timeprof}
Radial profiles of the time-scales studied in this work give further insights. In Fig.~\ref{F:profplot} we analyze time profiles as the azimuthal averages of \tdep\ and \torb\ for each Hubble type. The figure shows the radial variation of the molecular depletion time (\tdep, first row), orbital time (\torb, second row), and the ratio between the two time-scales (third row) within radial bins of 0.2\,R$_\mathrm{gal}$/\reff . In the fourth row, we plot the two time-scale azimuthal averages against each other.

The panel $a$ of the figure illustrates the behavior of the median of the depletion time for each category. The global average profile (the thick black line)  increases of about 0.5 dex from the center to 2\,$R_e$; the increase between one radius bin and the following is only 0.1 dex, though. A more significant gradient (1 dex from the centre to the outskirt of the galactic discs) was observed by \cite{wong2009} for a smaller sample of nearby star forming galaxies. \cite{utomo2017} show that depletion time increases, decreases, or does not change in the center of face-on galaxies in the EDGE sample\footnote{\cite{utomo2017} include only EDGE galaxies with inclination below $75^{\circ}$.}. Those behaviors seem to follow the trends of the \esfr\ profiles with the galactic radius. The authors attribute those changes to different dynamical pressure equilibriums between the galaxies induced by large dynamics (from barred and interacting systems) or local changes of stellar mass surface density. Molecular gas mass surface densities, instead, do not appear influenced by the same effects.

The orbital time profiles (Fig.~\ref{F:profplot}, panels $c$ and $d$) are obtained from our JAM modeling and are thus smooth, slowly increasing from the center to the outskirts of the galaxies. The average ratio between the two timescales is generally flat around $\sim1.5$ consistent with an orbital efficiency $\epsilon_\mathrm{orb}\sim5\%$.

Regarding the Hubble type, the molecular depletion times of early- and late-type galaxies are well separated in most of the cases (e.g. Fig.~\ref{F:profplot}$a$). The radial profiles of E, S0, and Sab galaxies are above the global average, while the radial profiles of Sb, Sbc, Scd galaxies are below the global average value. Nevertheless, the global average does not segregate early- and late-types at every radii. Sd galaxy profile shows an increase above the global average between $\sim0.7$ to $\sim1.3$\,$R_e$, where the depletion time assumes values very similar to early-type galaxies (i.e. \tdep$\sim10^{10}$\,yr). Scd galaxy profile displays a similar increament between $\sim1.3$ to $\sim2$\,$R_e$, where the depletion time is slightly below $\sim10^{11}$\,yr. For the earliest (E) and latest (Sd) types we possess only a few measurements (see histogram on the top of Fig.~\ref{F:profplot}), therefore their profiles might suffer of sample bias.

In the panel $b$, early-type galaxies show orbital times shorter than the sample average, while for late-type objects the times are longer than this value. In the panel $c$, the ratio of timescales appears completely segregated by morphology around the average sample profile.

The last panel of Fig.~\ref{F:profplot} shows the azimuthal averages of two time-scales against each other. In most of the cases, the radial increase of \torb\ is tracked by an increase in \tdep, while the Sd profile are a notable exception. The Sbc profile, in particular, follows closely the orbital efficiency of 5\%, meaning that this galaxy type drives the resolved relationship between \tdep\ and \torb\ . Indeed, Sbc galaxies dominate the total amount of detected lines-of-sight in our sample. In general, the correlations with the Hubble type that we see through this analysis closely resembles to the pixel-by-pixel behavior of the previous sections even when HI is accounted (see Appendix~\ref{A:hi}).

In summary, we observe that the galaxy morphologies are correlated to both the molecular depletion time and the orbital time of the galaxies. As a consequence, galaxies with different morphologies have different slopes in the \tdep-\torb\ relation. This behavior is mostly driven by \esfr, because \emol\ in our sample does not seem to correlate with either the orbital time or morphologies\footnote{Note that our measurements are done within the CO map masked region. This masked region does not encompass the whole galaxy and favors regions where CO can be detected.}. The observed segregations are not simply driven by the stellar mass, even though it has an important contribution in setting the molecular depletion time of the galaxies.

\begin{figure*}
\centering
{\includegraphics[width=\textwidth]{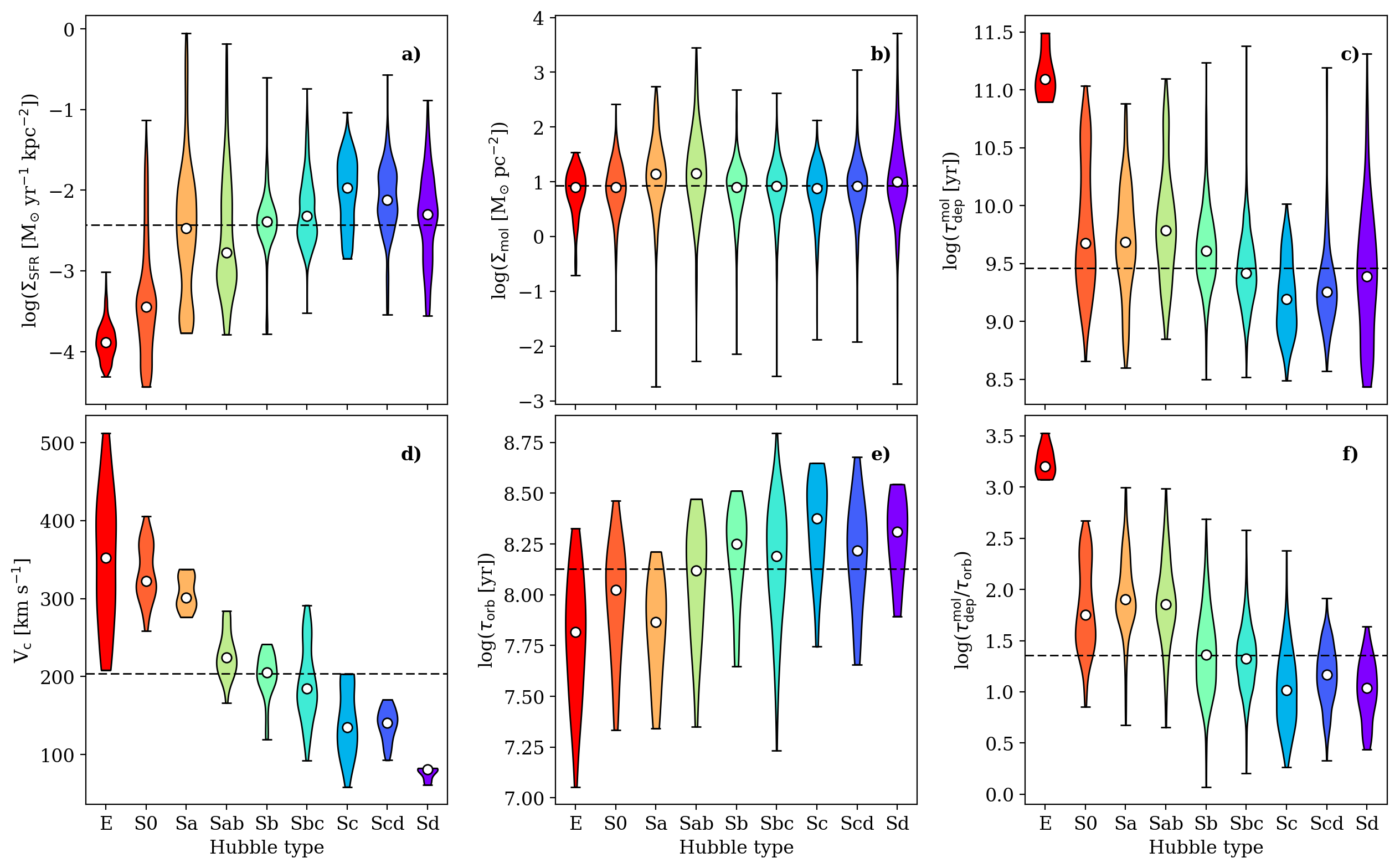}}
\caption{Violin plot representations of the resolved quantities that are studied in this paper across the Hubble types. Violin plots are histograms, where the width along the x-axis indicates the normalized fraction of data at the corresponding y-axis value. The values of \esfr, \emol, \tdep, and \tdep/\torb\ are analyzed pixel-by-pixel, while $V_c$ and \torb\ azimuthal averages are calculated within 2\,R$_\mathrm{eff}$. Choosing $V_c$ and \torb\ within 1\,R$_\mathrm{eff}$ renders the trend with Hubble type of the two quantities even more prominent. White circles represent the median values for each Hubble type. Dashed lines show the global median of the full sample of galaxies for a given property. This figure shows that all quantities, except \emol, appear to be related to the galactic morphology. These trends do not change if we include \emol\ and \esfr\ from the non-detected line-of-sights.}
\label{F:violins}
\end{figure*}

\section{Discussion}\label{S:discussion}
The \tdep-\torb\ relation is supposed to represent a scenario where both growth of gravitational instabilities, the subsequent collapse of gas clouds, and ultimately star formation are influenced by the galactic rotation. 

The existence of such a relation is not trivial, though. Star formation occurs exclusively within molecular clouds, which are local overdensities with respect to the molecular gas density distribution (for a recent review see \citealt{kennicutt_evans12}). Hence, \esfr\ and \emol\ can be considered as local quantities. The circular velocity, instead, depends on the mass within a certain galactocentric radius. The relationship between molecular depletion time and orbital time is, therefore, the parametrization of large-scale dynamics that acts on small scales.

\subsection{Morphological quenching in action?}\label{SS:disc_morphquench}
Perhaps, the most striking manifestation of large-scale dynamics is represented by galaxy morphology, historically classified via Hubble type. Several works have shown that the Hubble types correlated with many galactic integrated properties (see e.g, \citealt{roberts_haynes1994}; \citealt{consolandi2017}, and references therein). More recently the resolved study of \cite{gonzalez_delgado2015} observed that stellar metallicity, age, mass, and mass surface density increase monotonically from late- to early-type galaxies, while the opposite behavior apparent in their SFRs (\citealt{gonzalez_delgado2016}). In line with these works, in Fig.~\ref{F:violins}, we summarize the resolved galactic properties analyzed in this paper through morphology. On average, SFR surface density increases from early- to late-type morphologies, while the molecular depletion time decreases. The trend of \tdep\ with morphology has been noticed also by integrated studies (see, e.g. \citealt{saintonge2011}, their Fig. 5). Instead, the average behavior of the molecular gas surface density is generally flat with respect to the Hubble types. Previous studies have illustrated that early-type galaxies can encompass a large fraction of \htwo\ gas (e.g. \citealt{young2009}, \citealt{young2014}). Nevertheless, we remind the reader that our sample is IR-selected, therefore the early-type objects analyzed here might be more gas rich compared to ``standard'' E or S0 galaxies.

The existence of a flat behavior of \emol\ with respect to Hubble type might indicate that a significant reservoir of molecular gas is present in every galaxy of our sample, but in our early-types the cold gas has lost the ability to form stars (e.g., \citealt{martig2013}). The suppression of star formation in galaxies is generally referred as ``quenching''. Star formation quenching has been explained using a variety of effects (see the introduction of \citealt{martig2009} for a quick summary).

In particular, together with the conclusions of \cite{gonzalez_delgado2015}, our evidence  supports the idea of ``morphological quenching,'' where a gaseous disc embedded within a stellar spheroid, rather than a stellar disc, becomes stable against gravitational collapse (e.g., \citealt{martig2009}). This scheme does not envision a shortage of molecular gas, in line with what we observed in our data. Instead, the suppression of the disc instabilities is the cause of the star formation shutdown even if a substantial amount of gas is present. 

Other explanations can apply too. Quenching introduced by AGN feedback (e.g. \citealt{cattaneo2009}) might have an effect, since $\sim35$\% of our targets host an AGN. The discrepancy in the star formation rate along the Hubble sequence can be also enhanced due to the particular conditions within the late-type galaxies. In those systems, high mass star formation (mini-starburst events) can trigger feedback causing the destruction of the clouds, but also compression of the interstellar gas that creates new overdensities and increases star formation (\citealt{saintonge2011}). We can not exclude also the possibility that our early-type galaxies require a lower-than-Galactic $\alpha_\mathrm{CO}$ to deduce the right \emol\ values (see Appendix~\ref{A:caveats}).

\subsection{Stabilization via shear}\label{SS:disc_shear}

The morphological quenching model is based on the \cite{toomre64} theory, which predicts that the development of gravitational instabilities within the gaseous disc is hampered by the gas kinematics and by the presence of dissipative forces induced by the galaxy differential rotation, as shear. 

Shear as a stabilizing agent against self-gravity has been invoked several times to explain the differences of cloud properties between observations and simulations (e.g. \citealt{dobbs_pringle2013}, \citealt{colombo14a}, \citealt{suwannajak2014}, \citealt{miyamoto2015}, \citealt{ward2016}). \cite{hunter1998} defined a threshold in the gas mass surface density that sets a lower limit for the GMC formation which is directly proportional to the local shear rate. \cite{meidt2015} showed that in M51, the cloud lifetime appears to be constrained by the shear rate in the inter-arm regions of the galaxy where this effect dominates over stellar feedback. Several works indicated that shear does not only set the locations where clouds can form, but also control star formation itself. Hydrodynamical simulations of \cite{weidner2010} have shown that the formation of super star clusters is inversely proportional to the shear strength (see also \citealt{fogerty2016}). The likelihood of OB associations formation is also disfavored in this context. Similarly, \cite{hocuk_spaans2011} observed that clouds subjected to high shear from super-massive black holes tends to have lower SFE. \cite{davis2014} noticed some connection between shear strength and SFE in their sample of star forming early-type galaxies. Weak rotational support might also be one of the main causes of starbursts in high redshift systems. The low angular momentum of observed $z\approx 1-3$ objects is thought to reduce their stability and favor the generation of large clumps that substantially increases the star formation rate (see \citealt{obreschkow2015} and references therein). Nevertheless, star formation in the Milky Way clouds does not seems to correlate with shear at any stage of their evolution. \cite{dib2012} discussed that shear might have an effect in setting where GMCs can form, but self-gravity is mainly balanced by other factors as stellar feedback, turbulence or magnetic field. The sharp decrement of the mean circular speed across the Hubble type (Fig.~\ref{F:violins}$d$; see also \citealt{kalinova2017b}), together with the increment of the orbital time (Fig.~\ref{F:violins}$e$), suggests that a different degree of shear is present within the different morphologies.

\begin{figure*}
\centering
{\includegraphics[width=0.85\textwidth]{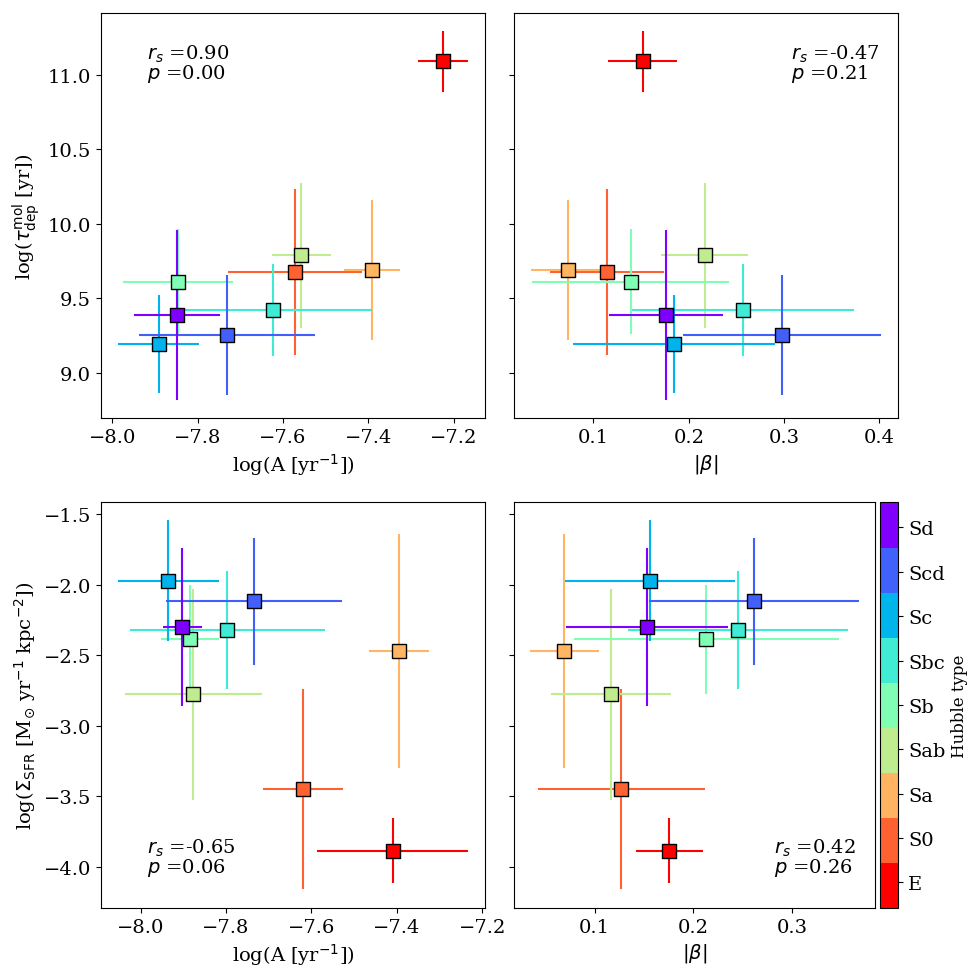}}
\caption{The correlations between the azimuthally-averaged shear rate ($A$, left), the absolute value of the rotation curve shape ($\beta$, right), the molecular depletion time (top), and the SFR surface density (bottom) across the Hubble types where both dependent and independent quantities are defined. Colored squares indicate the median of the quantity distributions within a given morphology, while the error bars show their median absolute deviations. In the corner of each panel, the Spearman rank ($r_s$) of each correlation is indicated together with the $p$-value ($p$). This figure shows that the average molecular depletion times and the SFR surface density are strongly correlated to the local shear across the Hubble types.}
\label{F:tdep_shear}
\end{figure*}

We can explicitly calculate the local shear rate within our galaxies through the Oort's constant $A$:

\begin{equation}\label{E:shear}
A = 0.5\left(\frac{V_c}{R_\mathrm{gal}} - \frac{dV_c}{dR_\mathrm{gal}}\right)=\frac{\pi}{\tau_\mathrm{orb}}(1-\beta),
\end{equation}

\noindent where $\beta=d\ln V_c/d\ln R_\mathrm{gal}$ represents the shape of the circular speed curve. Assuming that the rising part of the curve behaves as a solid body, $V_c\propto R_\mathrm{gal}$, this gives $\beta=1$ and $A=0$, while in the flat part of the curve, $V_c=$constant, $\beta=0$, and the shear reaches its maximum at $A=\pi/\tau_\mathrm{orb}$. In the top left panel of Fig.~\ref{F:tdep_shear}, we plot the average depletion time versus the average azimuthal shear ($A$) for each Hubble type. Clearly, the molecular depletion time follows the increment of $A$ across the Hubble types within the detected lines-of-sight, where the late-type galaxies have lower \tdep\ and less shear than the early-types. Moreover, the averaged \tdep\ and $A$ are strongly correlated across the Hubble types, showing a Spearman rank of $r=0.9$. This value decreases slightly to $r=0.86$ when the elliptical galaxies are removed, because these galaxies only have a few detections. A similar (but weaker) correlation between the shear rates and morphologies have been observed also through integrated measurements (\citealt{seigar2005}). The top-right panel of Fig.~\ref{F:tdep_shear} shows the azimuthal averaged values of $\beta$ for each galaxy in our sample. The average $\beta$ per morphology slightly decreases across the Hubble types ($r=-0.47$), which indicates again that the shear increases from late- to early-types. Generally, $|\beta|$ is always below unity in Fig.~\ref{F:tdep_shear} (right), meaning that, in the regions of the galaxies where we have detections, the contribution of the shear is never equal to zero. Therefore, according to Eq.~\ref{E:shear}, the shear behavior is dominated by the behavior of the orbital time in our sample. In Section~\ref{S:torb_hubmass} we observe that, on average, \esfr\ increases with the orbital time across the Hubble types (see Fig~\ref{F:hubmassplot}$b$). In light of Eq.~\ref{E:shear}, this suggests that \esfr\ is inversely proportional to the local shear as illustrated in Fig.~\ref{F:tdep_shear} (bottom-left).\\

\noindent Given this, there does not appear to be a universal ``Silk-Elmegreen'' law to appropriately describes the overall star formation in various morphologies of galaxies (at least on local scale), but a clear decrease of \tdep-\torb\ slope along the Hubble types is observed. Other parameters (such as shear) might need to be taken into account to obtain a universal star formation law (see also \citealt{tan2000}, \citealt{shi2011}, \citealt{krumholz2012}, \citealt{davis2014}, \citealt{utreras2016}, \citealt{bolatto2017}). Nevertheless the violin plots in Fig.~\ref{F:violins} show a large overlap, meaning that several regions in galaxies with different morphologies work in a similar way. The \tdep-\torb\ relation that we discussed in Section~\ref{S:res_main} might emerge due to the superposition of regions with similar local effects, but, not necessarily, belong to similar morphologies. In particular, spheroids and bulges are associated with discrepant data (Section~\ref{S:torb_hubmass} and Appendix~\ref{A:loc}), as well the discrepant points with high metallicity, stellar surface density, and velocity dispersion (as seen in Appendix~\ref{A:loc}).  From this coincidence, it may be that the discs follow the \tdep-\torb\ relation while bulges and spheroids do not. However, these results are strongly shaped by our sample selection, so that this interpretation is not robust.\\

\noindent In conclusion, it appears that galaxy morphology affects both the capability of galaxies to form stars and the galaxy dynamics in a broad sense. The reverse causality may also be true: dynamical and gravitational instabilities may have an impact on SFR that can change the galaxy's appearance through secular processes (\citealt{khochfar_silk2006}, \citealt{combes2009}). 

\section{Summary}\label{S:summary}
In this paper, we analyze the relation between molecular star formation effiency and large-scale dynamics on the plane described by molecular depletion time and orbital time through a sample of 39 EDGE-CALIFA kpc-resolved galaxies (with $i<65^{\circ}$) spanning various morphological types, SFRs, molecular gas masses, and stellar masses. Our findings are summarized below.

\begin{itemize}

\item Considering all our detected lines-of-sight, \tdep$\sim 20$\torb\ (i.e. $5\%$ of the available molecular gas is converted into stars at each orbit), with a large scatter of 0.5 dex (the left panel of Fig.~\ref{F:tdep_torb}). This result is in agreement with previous findings by \citet{kennicutt1998}, \citet{wong_blitz2002}, and \citet{leroy2013}. \\ 

\item The integrated measurements of molecular depletion time for Sb-Sbc type galaxies are moderately correlated with the orbital times at 2\,$R_\mathrm{eff}$ with a Spearman rank of $\sim 0.5 - 0.7$ (the right panel of Fig.~\ref{F:tdep_torb}). The early-type (E and S0) galaxies show very long depletion times, shifted away from the main \tdep$-$\torb\ relation. \\

\item Galaxies with different Hubble types appear to follow different \tdep-\torb\ resolved relations that decrease in steepness from the early- to late-types (Fig.~\ref{F:hubmassplot}).  Alternatively, the \esfr-\torb\ relation increases in steepness for later Hubble types. Those trends are less pronounced when binning the galaxies by their integrated stellar mass. On the other hand, the kpc-measurements of \emol\ do not correlate with the Hubble types, suggesting that different \tdep$-$\torb\ relations across the Hubble types are driven by \esfr, rather than \emol. However, our conclusion may be affected by the infrared-bright criterion in the sample selection of the EDGE survey. \\

\item The azimuthal averages of molecular depletion times become shorter from early-type to late-type galaxies, while the opposite behavior is observed for the orbital time (Fig.~\ref{F:profplot}). As a result, the ratios between \tdep\ and \torb\ decrease from early-type to late-type galaxies (the bottom right panel of Fig.~\ref{F:violins}). \\

\item On average, the local shear rate appears to decrease across the Hubble types and appears to correlate with the molecular depletion times (Fig.~\ref{F:tdep_shear}), with a Spearman rank correlation coefficient of 0.9. This result provides a tentative evidence for a scenario where shear plays an important role in counteracting gravitational contraction and possibly suppressing star formation.\\

\end{itemize}

\noindent This study highlights also the urgency to gather a more homogeneous sample (in term of both galaxy morphologies and stellar masses) of kpc-resolved observations, together with the observations of molecular gas and optical IFU data at the scale of molecular clouds ($\sim 50$ pc), thereby connecting the global and local effects in a more consistent way.

%=====================================================================
% ACKNOWLEDGMENTS
%=====================================================================

\section*{Acknowledgements}
The authors thank the anonymous referee for the useful insights that largely improved the quality of the paper. DC thanks Axel Weiss and Sharon Meidt for the stimulating discussions. DC acknowledges support by the Deutsche Forschungsgemeinschaft, DFG through project number SFB956C. The works of DU and LB are supported by the National Science Foundation (NSF) under grants AST-1140063 and AST-1616924. ADB and RCL acknowledge support from NSF through grants AST- 1412419 and AST-1615960. ADB also acknowledges visiting support by the Alexander von Humboldt Foundation. TW acknowledges support from NSF through grants AST- 1139950 and AST-1616199. SFS acknowledges the PAPIIT-DGAPA-IA101217 project and CONACYT-IA- 180125. ER is supported by a Discovery Grant from NSERC of Canada. SV acknowledges support from NSF AST- 1615960. HD acknowledges financial support from the Spanish Ministry of Economy and Competitiveness (MINECO) under the 2014 Ramón y Cajal program MINECO RYC-2014-15686. We acknowledge the usage of the HyperLeda database (http://leda.univ-lyon1.fr). Support for the CARMA construction was derived from the states of California, Illinois, and Maryland, the James S. McDonnell Foundation, the Gordon and Betty Moore Foundation, the Kenneth T. and Eileen L. Norris Foundation, the University of Chicago, the Associates of the California Institute of Technology, and NSF. This research is based on observations collected at the Centro Astronomico Hispano Aleman (CAHA) at Calar Alto, operated jointly by the Max-Planck Institute for Astronomy (MPIA) and the Instituto de Astrofisica de Andalucia (CSIC). This research made use of Astropy, a community-developed core Python package for Astronomy (\citealt{astropy2013}).

%=====================================================================
% REFERENCES
%==============================BIBTEX=======================================
\footnotesize{
\bibliographystyle{mn2e_new}
\bibliography{cold}
}

%=====================================================================
% APPENDICES
%=====================================================================
\appendix

\section{Testing the Effects of Non-detections}\label{A:nondetections}
\begin{figure*}
\centering
{\includegraphics[width=1\textwidth]{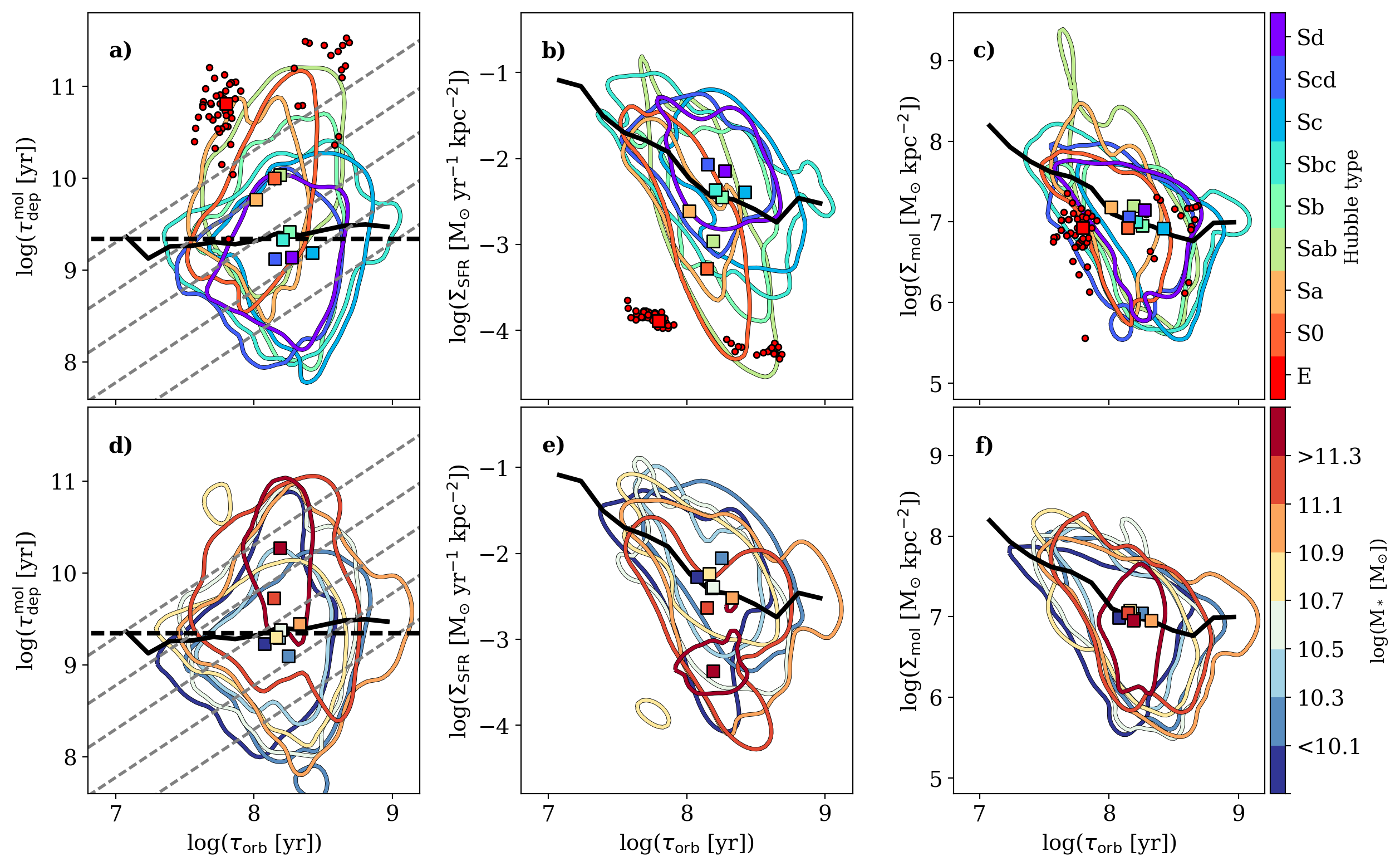}}
\caption{Pixel-by-pixel relations between \tdep\ (first column), \esfr\ (second column), and \emol\ (third column) with respect to \torb\ color-encoded with Hubble type (first row) and integrated stellar mass (second row) where molecular depletion time non-detections are added. Symbols and notations follow Fig.~\ref{F:hubmassplot}. \emol\ non-detections are the most significant. The increased scatter in the \tdep-\torb\ relationship, however, does not change the main conclusion of the paper regarding the role of morphology in the dynamics and star formation properties of the galaxies.}
\label{F:hubmassplot_nondet}
\end{figure*}

In Section~\ref{S:torb_hubmass} we observed that, binned through their morphology, \tdep\ and \torb\ line-of-sight averages are anti-correlated, while \esfr\ and \torb\ correlate. Instead \emol\ does not show any significant trend with both orbital time or Hubble type. To understand how non-detections could affect those behaviors, we reproduce Fig.~\ref{F:hubmassplot} including \esfr\ and \emol\ lower limits. The result is shown in Fig.~\ref{F:hubmassplot_nondet}. In Section~\ref{S:res_main} we observed, for example, that the addition of the depletion time non-detections yielded no discernible correlation between \tdep\ and \torb\ when observed pixel-by-pixel.

Mainly the non-detections affect the \tdep\ and \emol\ trends of late-type galaxies (from Sb to Sd) for which relation contours extend to lower times and surface densities, respectively (Fig~\ref{F:hubmassplot_nondet} panels $a$ and $c$). Elliptical galaxies show several new data points at longer orbital times. In this representation, the late- type galaxies do not show a clear \tdep-\torb\ relation. The \esfr-\torb\ diagram is not largely affected by those values, though (Fig~\ref{F:hubmassplot_nondet}$b$). Together, the behavior of the early-type galaxies remains significantly different from the late-types and the average behaviors observed through detections only are largely preserved.

In a similar way, the inclusion of non-detections does not alter our conclusions about the role of the stellar mass in the trends noticed through detections only via Hubble type. In Section~\ref{S:torb_hubmass} we observed a weak (anti-) correlation between the average (\esfr) \tdep\ and \torb . In Figure~\ref{F:hubmassplot_nondet} (second row) those behaviors look less defined. As for the galaxy morphologies, we do not distinguish any clear trends when \emol\ is considered (Fig.~\ref{F:hubmassplot_nondet}$f$).

\section{Testing the Influence of Local Galactic Properties}\label{A:loc}

In this paper, we use the nebular emission lines surveyed by CALIFA to obtain information about the star formation in the galaxies. CALIFA also observed the stellar spectra, which can be used to characterize the stellar population properties, including the mass surface densities, metallicities, ages, and stellar kinematics (see \citealt{perez2013}; \citealt{cid_fernandes2013}; \citealt{cid_fernandes2014}; and \citealt{gonzalez_delgado2014a} for further details). Here we study how the relation between the molecular depletion time and the orbital time is influenced by local properties. To do so, we approach the problem on two fronts: by considering (1) the properties of the stellar population, and (2) the kinematics within galaxies. For this study, we consider the detections only, because the contours related to detections plus non-detections provide similar conclusions.

\begin{figure*}
\centering
{\includegraphics[width=\textwidth]{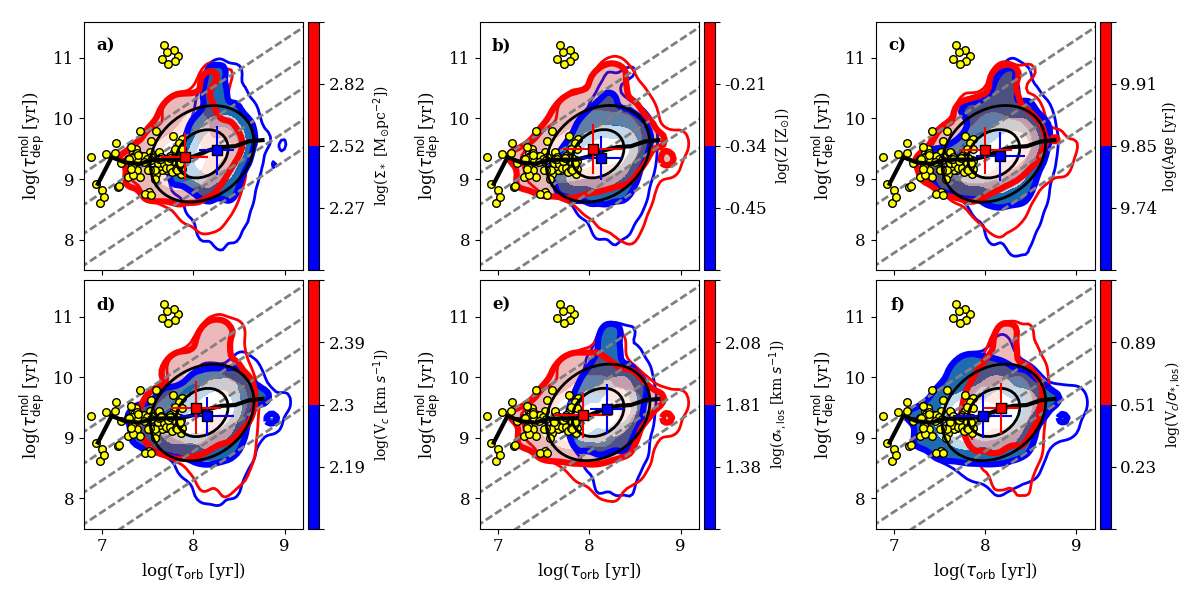}}
\caption{Analysis of the distribution of local stellar properties in the \tdep-\torb\ diagram: stellar mass surface density (panel $a$), metallicity (panel $b$), age (panel $c$), circular speed (panel $d$), line-of-sight stellar velocity dispersion (panel $e$), and line-of-sight stellar order-over-random motion (panel $f$). The data points are separated according to their color-encoding property medians. The contour shades contain 50\%, 75\%, and 95\% of the total number of detections. The outer thick (thin) colored contour contains 95\% of the detections (detections plus non-detections), while the colored squares show the median of \tdep\ and \torb\ within the respective distributions with their respective standard deviations (for the detection only). The black solid lines indicate median of the related quantities within bins of 0.2 dex with respect to \torb . The confidence ellipsoids shown in black contain $\sim68\%$ and $\sim95\%$ of the total number of detections. Gray dashed diagonal lines indicate constant orbital efficiency ($\epsilon_\mathrm{orb}$): from the top to bottom 0.5\%, 1.7\%, 5\%, 17\%, and 50\%. Yellow circles indicate the measurements obtained within the bulge area delimited by the bulge effective radius (Neumann et al. 2017). The figure shows that the scatter is mainly given by data points with high values of \estar, stellar metallicity, and velocity dispersion generally attributable to nuclei, bulges, or spheroids.}
\label{F:ttpropplot}
\end{figure*}

Following Bolatto et al. (2017; their Fig. 17 and 18), we divide the sample into two subsamples based on the properties under study. The results of our analysis are illustrated in Fig.~\ref{F:ttpropplot}. We plot in blue (red) the data with values below (above) the property median. We also use yellow circles to show measurements obtained within the bulge area as described by the effective radius calculated by \cite{neumann2017}. In general, the overlapping region between the upper and lower hand distributions contains the majority of the points ($90-95\%$ of the total). However, outliers also give useful insights.

The stellar mass surface density is the local property that appears to have the most direct influence on both depletion and orbital times (Fig.~\ref{F:ttpropplot}$a$). The medians of the lower and upper hand distributions (red and blue squares in the panel) closely follow the running average between \tdep\ and \torb\ (full black line in the panel). Lower orbital times are dominated by high values of \estar. The converse is also true when we consider the non-detections (thin colored lines in the plot). The data are not, however, separate across the lines of constant depletion time. This means that both star formation rate and molecular mass surface densities are correlated with \estar . The stellar mass surface density is also partially connected with the orbital efficiency: low efficiencies appear dominated by high \estar\ values and vice versa. The outliers of the relation (i.e. the data points outside the outer confidence ellipsoid) show mainly anomalous \estar\ values. The stellar mass surface density represents the local gravitational potential, therefore a connection with \torb, which in turn parametrizes the global potential at a particular galactocentric radius, is expected.  Several authors (e.g., \citealt{leroy2008}, \citealt{bolatto2017}) have also noticed that the depletion time anti-correlates with \estar . This is foreseen by a picture of feedback-regulated star formation (e.g., \citealt{ostriker_shetty2011}), where the dynamical-equilibrium pressure is correlated with \esfr . The molecular-to-atomic to gas ratio is also proportional to the hydrostatic pressure in the disc mid-plane (e.g, \citealt{wong_blitz2002}, \citealt{blitz_rosolowsky2006}), and this quantity is covariant with the stellar mass surface density (\citealt{kim2013}).

Fig.~\ref{F:ttpropplot}$b$ partitions the \torb-\tdep\ data by metallicity.  Again, most data overlap between the populations, but the outliers are associated with high metallicity, specifically high \tdep\ and low \torb\ , and (considering the non-detections also) the high orbital time data all show metallicity values in the high metallicity group. Interestingly, the detections with a metallicity in the lower half of the distribution are tightly constrained within the 95\% confidence ellipsoid. The lower orbital efficiency shown in the plot ($\epsilon_\mathrm{orb}=0.5\%$) is fully dominated by high metallicity values. Indeed, the median related to the high metallicity distribution  seems slightly shifted towards low orbital efficiencies and vice versa. The same is true if we consider the behavior of the stellar age across the relation (Fig.~\ref{F:ttpropplot}$c$). In this case, however, the contours that contain 95\% of the data for both distributions appear co-spatial in the diagram. If we include non-detection too, the lines-of-sight belonging to the lower age data extend to lower depletion time values.

Stellar kinematics give further insights. Interestingly, different values of circular speed are distributed across all different values of orbital times (Fig.~\ref{F:ttpropplot}$d$), meaning that the orbital time per pixel is mainly driven by the galactocentric radius. Depletion times above $10^{10}$\,yr are located only towards high $V_c$ values. This result reflects what observed in Bolatto et al. (2017; their Fig. 18), where high stellar mass lines-of-sight dominate high \tdep\ region in their plots. The medians of \torb\ and \tdep\ within the two distributions closely resemble the medians when the relation is color-encoded by stellar metallicity or age. 

Instead the behavior of the stellar velocity dispersion ($\sigma_*$) is similar to the behavior of \estar\ (Fig.~\ref{F:ttpropplot}$e$). Again outliers appear dominated by high velocity dispersion lines-of-sight. The trend of the ordered-over-random motion of the stars mirrors the $\sigma_*$ behavior (Fig.~\ref{F:ttpropplot}$f$) and it is clearly driven by the velocity dispersion.

Lines-of-sight within the identified galactic bulge areas (\citealt{neumann2017}) are primarily found as outliers in these diagrams, particularly within the upper distributions of stellar mass surface density, metallicity, and velocity dispersion.

In conclusion, it appears that outliers of the \tdep-\torb\ relation are generally dominated by high values of \estar, $\sigma_*$, and, in particular, stellar metallicity. These properties are associated with galaxy nuclei, bulges, and spheroids. These outliers are also found in data from early-type galaxies, where these structures are dominant (see Fig~\ref{F:hubmassplot}$a$). Stellar surface density and velocity dispersion closely follow the trend of the relation, while we have some indications that high metallicity, ages, and circular speeds belong to lines-of-sight shifted towards low $\epsilon_\mathrm{orb}$ and vice versa.

\section{Testing the Influence of Atomic Gas}\label{A:hi}

\begin{figure*}
\centering
{\includegraphics[width=\textwidth]{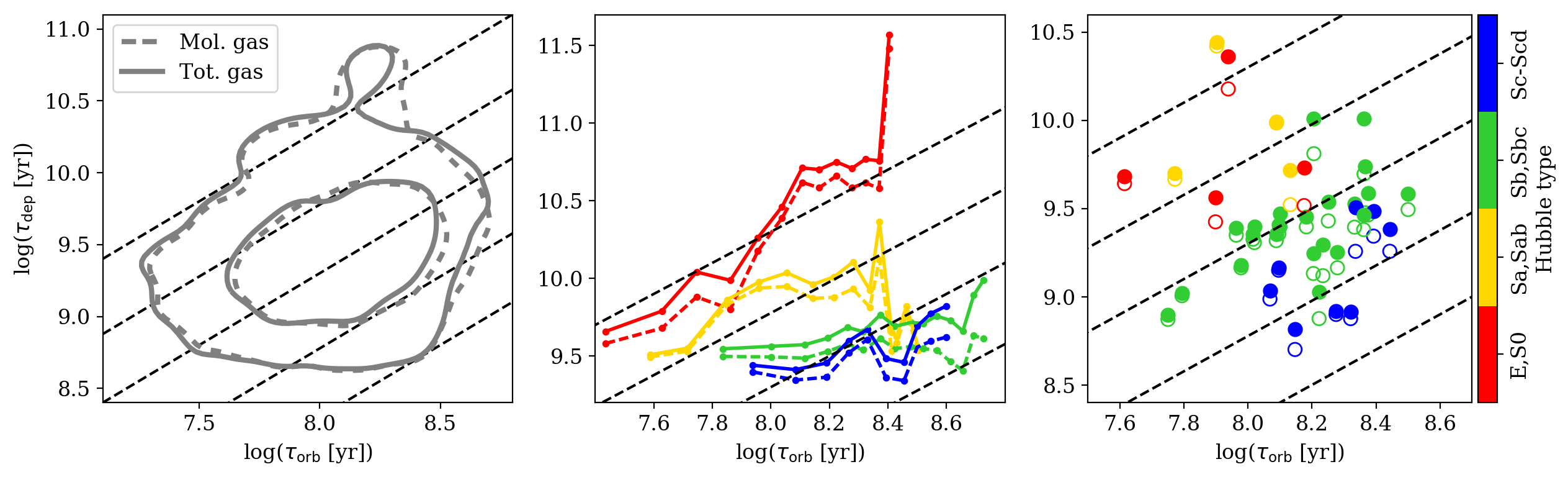}}
\caption{The depletion time - orbital time relations explored in the paper with synthetic atomic gas measurements added. \emph{Left:} pixel-by-pixel $\tau_{\rm dep}$-\torb\ relations from total (atomic plus molecular; gray full lines) and molecular (gray dashed lines) gas. The inner and outer contours contain 66\% and 95\% of the total detected CO lines-of-sight, respectively. \emph{Middle:} azimuthal averages of $\tau_{\rm dep}$ and \torb\ in radial bins of 0.2\,$R_{\rm gal}/R_e$ where total gas (full lines) and molecular gas only (dashed lines) are included in the calculation of $\tau_{\rm dep}$, divided into Hubble type groups. The last group, ``Sc-Sd'', considers ``Scd'' galaxies too. \emph{Right:} total (full symbols) and molecular (empty symbols) gas depletion times integrated within 2\reff\ with respect to orbital time measurements at 2\reff . Each symbol represents a given galaxy color-encoded according to its Hubble type group. In the panels, dashed black straight lines indicate constant conversion efficiencies, from top to bottom, of 0.5\%, 1.7\%, 5\%, 17\%, and 50\%. The addition of the atomic gas does not appear to significantly alter the main conclusions of our analysis of the \tdep\-\torb\ relationship.}
\label{F:hitest}
\end{figure*}

Kennicutt-Schmidt and Silk-Elmegreen relations have been originally measured for the total neutral gas which incorporate both molecular and atomic gas, the latter generally traced via 21\,cm wavelength line observations.  In this Section, we test how the inclusion of the atomic gas would potentially change the Silk-Elmegreen type relation that we analyzed in this paper. Since the resolved HI measurements do not exist for our sample, we generate synthetic atomic mass surface density by imposing $\Sigma_{\rm HI}=10$\,M$_{\odot}$\,pc$^{-2}$ for each pixel, corrected for the physical size of the pixel. This is justified by previous results. For example, \cite{leroy2008} showed that the radial profiles of atomic mass surface density are remarkably constant around $10$\,M$_{\odot}$\,pc$^{-2}$ (see their Appendix F) in a sample of 23 nearby star forming galaxies drawn from THINGS (\citealt{walter08}) and HERACLES (\citealt{leroy2009}). Fig.~\ref{F:hitest} shows the result of the test, which considers detections only. Generally, the addition of the atomic gas does not change the appearance of pixel-by-pixel relation originally observed in Fig~\ref{F:tdep_torb} (see Fig.~\ref{F:hitest}, left). The inner contours of the relations that include 66\% of data points from total and molecular gas only (full and dashed lines, respectively) follow the 5\% orbital efficiency line. This might be related to the fact that most of our galaxies appear to be molecular-dominated, since in 75\% of our pixels $\Sigma_{\rm mol} > \Sigma_{\rm HI}$ across all sampled galacto-centric radii. The middle panel of Fig.~\ref{F:hitest} shows the azimuthal average profiles of the total (full lines) and molecular (dashed lines) gas depletion time profiles with respect to the orbital time profiles divided into Hubble type groups. Total and molecular-only gas profiles track each other quite well in each morphology group, with the total gas profiles always shifted toward slightly longer depletion times. Integrated quantities in the rightmost panel of the figure mirror this behavior. The anti-correlation with orbital times across the Hubble sequence is present whether total or molecular gas depletion times are considered. Nevertheless, vertical shift between the two depletion times is more prominent in some galaxies than in others and does not necessarily follows the morphological type. Summarizing, the conclusions concerning the correlation between galactic morphology, orbital time, and depletion time discussed along the paper are preserved by including the atomic gas contribution in the analysis, despite the representation used. However, since atomic gas-dominated surface densities do not correlate with \esfr\ (e.g. \citealt{schruba11}), the relation between total gas depletion time and orbital time might be associated to the conversion between HI to H$_2$ rather than to star formation itself. In particular the creation of molecular gas overdensities from the smooth atomic medium in the regions of the galaxies (where the gas self-gravity overcomes tidal forces and shear) might happen on the orbital timescale. Nevertheless, we are not able to discern between the two different interpretations with the current data.

\section{Caveats}\label{A:caveats}
\noindent \emph{Sample biases}. EDGE was designed to span larger values of stellar masses and morphologies than previous surveys. Nevertheless, the CARMA sample was selected based on IR-brightness, which leads to a lack of well-resolved E and Sd types, as well as galaxies with stellar mass below $10^9$\,$M_\odot$ and above $10^{11}$\,$M_\odot$. We further restricted the sample to galaxies with inclination  $<65^{\circ}$ and with good dynamical models. We ended up with a sample of 39 galaxies and 6360 individual lines-of-sight, each representing a $\sim$kpc-scale region. Of these, $80\%$ belong to Sab-to-Sc morphologies, and the remaining $20\%$ are equally shared between the other Hubble types. The percentage of Sab-to-Sc data points decreases to $75\%$ if non-detections are considered. At the same time, galaxies with stellar masses between 10$^{10}$-10$^{11}$\,M$_{\odot}$ dominate the data ($74\%$), while we have only few measurements for low-mass galaxies ($6\%$ for $M_*<10^{10}$\,M$_{\odot}$) and for high mass galaxies ($20\%$ for $M_*>10^{11}$\,M$_{\odot}$).  However, we find that these results do not change if we consider non-detections and that the sample used in this paper is representative of the inclination-limited EDGE sample (see Section~\ref{SS:sample}).\\

\noindent \emph{Flux recovery}. EDGE cubes do not include the total power data. The observing strategy, which incorporates data from CARMA's D- and E-configurations, allows the recovery of a large range of scales (see \citealt{bolatto2017}).  However, we cannot be sure about how the addition of the total power might influence the non-detections. Along the paper we observe that including the lower limits (especially of \emol\ ) can change the results we derive. In particular, CO non-detections add shorter depletion times to our measurements which results in no discernible correlation between \tdep\ and \torb\ . Total power data might boost \emol\ non-detections to larger values which would reestablish the relationship we observe through detections only. Therefore, our results regarding the \tdep-\torb\ relation and its variation with respect to the morphology of the galaxies need to be verified with a more homogeneous sample of targets and a more complete reconstruction of emission from the interferometer data.\\ 

\begin{figure}
\centering
{\includegraphics[width=0.45\textwidth]{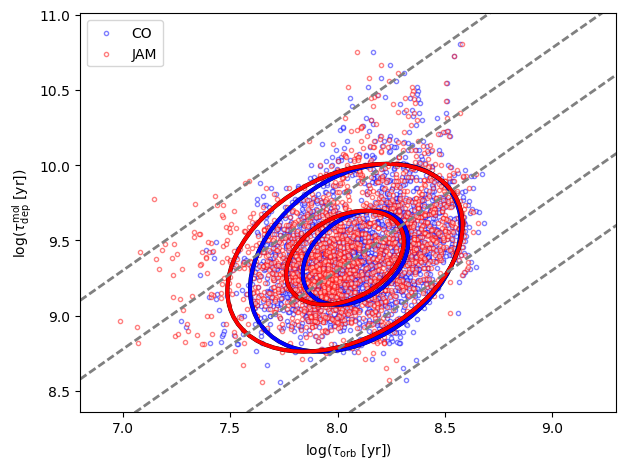}}
\caption{Pixel-by-pixel \tdep-\torb\ relations where orbital time is obtained with the JAM dynamical model (red) and CO rotation curves (blue). Dashed lines and confidence ellipses follow the convention of Fig.~\ref{F:tdep_torb}. The choice of rotation curve modeling does not significantly change the results.}
\label{F:tdep_tdyn_comp}
\end{figure}

\noindent \emph{CO-to-H$_2$ conversion factor variation}. For our sample, we derive the molecular gas mass surface densities from CO luminosities using a constant, Galactic conversion factor: $\alpha_\mathrm{CO}=4.4$\,M$_{\odot}$\,pc$^{-2}$\,(K\,km/s\,pc$^2$)$^{-1}$. Bolatto, Wolfire \& Leroy (2013; see also \citealt{accurso2017}) discuss that one of the main sources of influence of the CO-to-H$_2$ conversion factor is the metallicity. Low metallicity reduces the \htwo\ self-shielding that, in turn, decreases the quantity of available CO. Therefore, low metallicity galaxies require a larger $\alpha_{\mathrm{CO}}$ to derive a correct molecular gas mass from the observed CO emission. The global gas-phase metallicities of our galaxies are almost indistinguishable from the Solar value of 12 + $\log$[O/H] = 8.7. Nevertheless, this effect may be important for the local measurements that show values significantly lower than the Galactic average (see, e.g, Fig.~\ref{F:ttpropplot}) and might require larger values of $\alpha_\mathrm{CO}$ to obtain the right \emol . We do not believe that this bias dominates our results because the lowest gas-phase metallicity data points do not correspond to the lowest \emol\ values.

The usage of a constant $\alpha_\mathrm{CO}$ might not be appropriate for all galaxy morphologies. Massive galaxies tend to have stronger interstellar radiation field and higher gas velocity dispersion than low mass ones, which would increase the CO emission we measure. Those galaxies would require a lower-than-Galactic $\alpha_\mathrm{CO}$ to produce the correct amount of molecular gas mass. This would result in lower values of \tdep\ for this kind of galaxies. Considering Fig.~\ref{F:violins}, in order to have, in early-type galaxies, molecular depletion times comparable to the global average we would need an $\alpha_\mathrm{CO}$ $\sim2-3$ times lower than Galactic for Sa and S0 galaxies, respectively. We do not consider elliptical galaxies since we have too few detections for a robust conclusion. Here, we have taken the Milky Way as a Sbc galaxy (\citealt{gerhard2002}), where Sbc galaxies dominate our \tdep\ sample average. Given an uncertainty of $\sim0.3$\,dex on the generally assumed $\alpha_\mathrm{CO}$ (\citealt{bolatto2013}), it is plausible that the flat trend of \emol\ we observe with respect to Hubble type is caused by our choice of a constant CO-to-H$_2$ conversion factor.

Another aspect, discussed in \cite{leroy2013} and \cite{sandstrom2013}, is the possibility that galaxy centers have lower $\alpha_\mathrm{CO}$. This would lead to lowering the value of the depletion time at small galactocentric radii. At the same time, galaxy centers tend to have shorter orbital times. In this sense, by using a customized $\alpha_\mathrm{CO}$ in particular regions of the galaxies would result in a better correlation between \tdep\ and \torb\ as observed in Leroy et al. (2013, their section 4.4).\\

\begin{figure*}
\centering
{\includegraphics[width=\textwidth]{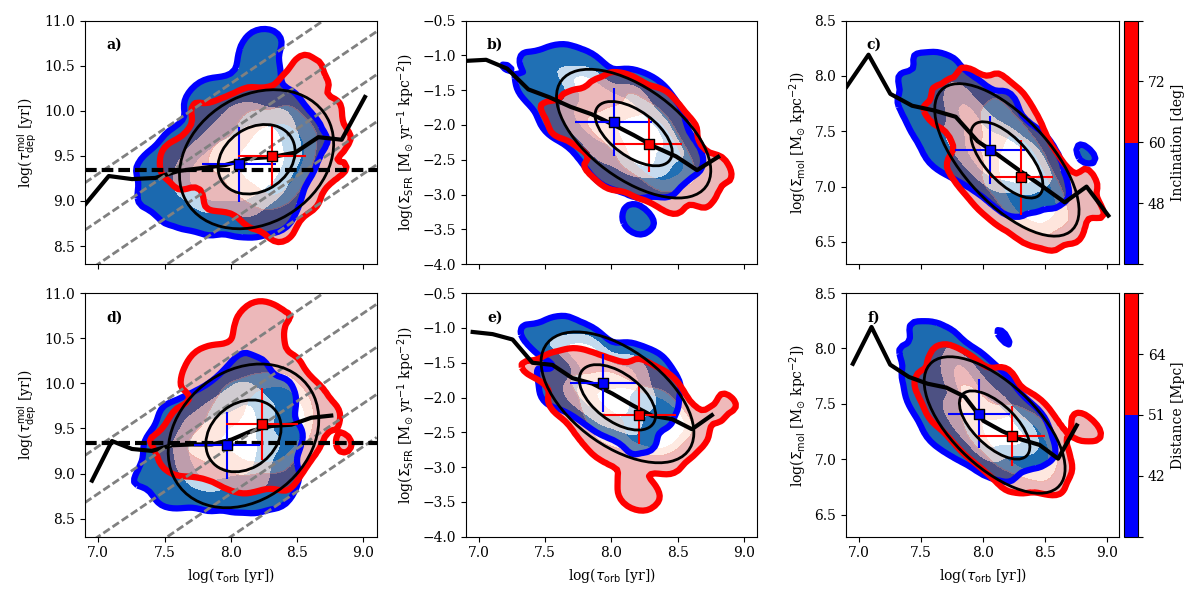}}
\caption{Relationships between \tdep, \esfr, and \emol\ with respect to the orbital time color-encoded by inclination (first row) and distance (second row). Symbols and notations follow Fig.~\ref{F:ttpropplot}.}
\label{F:caveats}
\end{figure*}

\noindent \emph{JAM versus CO rotation curves}. To test how the usage of dynamical models instead of CO rotation curves might influence our conclusions about the \tdep-\torb\ relation, we collect 13 galaxies (2234 data) with both dynamical models and well defined CO rotation curves, and we derive the orbital times for both of them. For this experiment we use only detections. CO rotation curve models have been calculated via the universal rotation curve formula presented in \cite{persic1996} (R.C. Levy et al., in preparation for more details). In Fig.~\ref{F:tdep_tdyn_comp} we plot the \tdep-\torb\ relation obtained through JAM- and CO-derived circular velocity curves. Confidence ellipses related to JAM appear shifted toward lower orbital times by $\sim$0.1 dex. This might be because the JAM approach tends to overestimate the mass-to-light ratio, therefore the circular velocity, in the center of the galaxies. This has been reported in several occasions: face-on barred systems (\citealt{lablanche2012}) or when averaging several stellar populations (\citealt{davis2013}). At the same time, CO rotation curves are severely affected by beam smearing at lower galactocentric radii, which underestimates velocities in those regions. Nevertheless, the same relationship between molecular depletion time and orbital time emerges from the CO-based rotation curves, so we conclude that this choice does not strongly affect our results. Most of the data overlap and the relations derived by the two modeling approaches are well described by the $5\%$ efficiency line with a 0.5 dex scatter.\\

\noindent \emph{Inclination and distance}. Some of the trends we observe may be influenced by galactic biases. To check this, in the first and second rows of Fig.~\ref{F:caveats}, we plot the SE law data color-encoded by galactic inclination and distance, respectively. For this test we use only detections.

For the inclination test, we include the full sample of galaxies at any inclination (where this work so far has only considered $i<65^{\circ}$). High orbital times are dominated by highly inclined galaxies. The same is true for data with very low \htwo\ mass and SF surface densities, besides the deprojection of the surface density for the inclination (see \citealt{utomo2017}). Depletion time seems less dominated by this parameter. However, the highest \esfr\ and \emol\ values are found in low inclination galaxies.

Although CALIFA has been designed as a diameter-limited survey to reduce the effect of the distance biases, the distance to the galaxies nonetheless appears to have some impact on the relations. The more distant objects dominate very high depletion time measurements ($>10^{10}$\,yr). \esfr\ and \emol\ appear shifted toward lower values. Orbital times above $10^{8.5}$\,yr, which constitute the prominent outliers, are located in the same bin. For the closest objects, instead, we can measure some of the lowest depletion time values, and the highest SFRs and \htwo\ masses per unit area of the sample. From the inclination and distance tests we conclude that the detection criteria select different data for different galaxy types and environments. The scaling with distance suggests that we may be detecting a small scale-dependent effect at the kpc-scales, but our data lack the spatial dynamic range to explore this extensively. These tests suggest small shifts in the typical galaxies in each population, however, the fundamental relationships still show the same basic trends no matter how the data are divided or selected.

%=====================================================================
% END DOCUMENT
%=====================================================================

\bsp % ``This paper has been produced using the ...''

\label{lastpage}

\end{document}